\documentclass[journal,twocolumn]{IEEEtran}
\usepackage[cmex10]{amsmath}
\usepackage{amssymb}
\usepackage{cite}
\usepackage{graphicx}
\usepackage{array,color}
\usepackage{multirow}
\usepackage{amsmath}
\usepackage{stfloats}
\usepackage{graphicx}
\usepackage{subfigure}
\usepackage{tabularx}
\usepackage{epsfig,epsf,color,balance,cite}
\usepackage{setspace}
\usepackage{bm}
\usepackage{textcomp}
\usepackage{algorithmic}
\usepackage{algorithm}
\usepackage{caption} 

\usepackage{comment}
\usepackage{subfigure}

\usepackage{caption}
\usepackage{graphicx}
\usepackage{setspace}
\usepackage{diagbox}

\usepackage{multirow}
\usepackage{graphicx}
\makeatletter

\newcommand{\Rmnum}[1]{\expandafter\@slowromancap\romannumeral #1@}
\makeatother
\usepackage{booktabs}
\usepackage{amsmath}

\makeatother
\pdfoutput=1
\begin{document}

\title{NMBEnet: Efficient Near-field mmWave Beam Training for Multiuser OFDM Systems Using Sub-6 GHz Pilots}
\author{
Wang Liu, Cunhua Pan, $\textit{Senior Member, IEEE}$, Hong Ren, $\textit{Member, IEEE}$, Cheng-Xiang Wang, $\textit{Fellow, IEEE}$, Jiangzhou Wang, $\textit{Fellow, IEEE}$, and Xiaohu You, \textit{Fellow, IEEE}\thanks{
\emph{Corresponding author: Cunhua Pan and Hong Ren.}

Wang Liu, Cunhua Pan, Hong Ren and Xiaohu You are with National Mobile Communications Research Laboratory, Southeast University, Nanjing 210096, China. (e-mail: {w\_liu, cpan, hren, xhyu}@seu.edu.cn).

Cheng-Xiang Wang is with the National Mobile Communications Research Laboratory, School of Information Science and Engineering, Southeast University, Nanjing 210096, China, and also with the Purple Mountain Laboratories, Nanjing 211111, China (e-mail: chxwang@seu.edu.cn).

Jiangzhou Wang is with the School of Engineering, University of Kent, CT2 7NZ Canterbury, U.K. (e-mail: j.z.wang@kent.ac.uk).

%
		
 } }

\maketitle

\begin{abstract}
Combining millimetre-wave (mmWave) communications with an extremely large-scale antenna array (ELAA) presents a promising avenue for meeting the spectral efficiency demands of the future sixth generation (6G) mobile communications. This technology achieves a high data rate and establishes high-gain directional transmission links. However, beam training for mmWave ELAA systems is challenged by excessive pilot overheads as well as insufficient accuracy, as the huge near-field codebook has to be accounted for. In this paper, inspired by the similarity between far-field sub-6 GHz channels and near-field mmWave channels, we propose to leverage sub-6 GHz uplink pilot signals to directly estimate the optimal near-field mmWave codeword, which aims to reduce pilot overhead and bypass the channel estimation. Moreover, we adopt deep learning to perform this dual mapping function, i.e., sub-6 GHz to mmWave, far-field to near-field, and a novel neural network structure called NMBEnet is designed to enhance the precision of beam training. Specifically, when considering the orthogonal frequency division multiplexing (OFDM) communication scenarios with high user density, correlations arise both between signals from different users and between signals from different subcarriers. Accordingly, the convolutional neural network (CNN) module and graph neural network (GNN) module included in the proposed NMBEnet can leverage these two correlations to further enhance the precision of beam training. To better evaluate the performance of the proposed algorithm, we employ the state-of-the-art system simulation software to obtain realistic channel data. Simulation results demonstrate the superior performance of the proposed strategy compared to the exhaustive search scheme and existing deep learning-based schemes.

\end{abstract}

\vspace{-0.4cm}
\section{Introduction}
Since millimetre-wave (mmWave) bands can offer abundant spectral resources and support high transmission rates, mmWave communications are pivotal in both current fifth generation (5G) and anticipated sixth generation (6G) mobile networks. \cite{chengxiang_wang, mmwave_1}. However, mmWave signals would experience greater path loss during propagation compared to sub-6 GHz signals. To mitigate path loss, 5G communication systems deploy massive antenna arrays at the base station (BS) for establishing directional transmissions. \cite{5g_mimo_1,5g_mimo_2,5g_mimo_3}. For 6G communication systems, the BS is expected to deploy more antennas and extremely large-scale antenna arrays (ELAA) have been proposed in\cite{xlmimo}. Benefiting from more antennas, ELAA can achieve higher directional transmissions gain, greatly improving the spectral efficiency of the communication system. \cite{cui_1}. Hence, combining mmWave communications with ELAA represents a promising technology to meet the rapidly growing demand for spectral efficiency in future 6G communication systems. \cite{mmwave_elaa}.

In mmWave communication systems, enhancing received signal power involves widely employing beam training based on a predefined codebook. Here, the BS searches for the codeword with the highest gain in the predefined codebook to form a directional beam. However, beam training is challenging for communication systems with ELAA. Specifically, the deployment of ELAA leads to an expansion of the boundary between the near field and the far field, i.e., the Rayleigh distance, which is proportional to the size of the antenna array \cite{def_rayleigh}. Hence, it's crucial to consider near-field communication based on the spherical wave assumption, since users are often located within the near-field domain. In near-field communications, specialized near-field codebooks tailored for spherical waves must be adopted. These codebooks contain significantly more codewords than traditional far-field codebooks because near-field codewords need to be searched not only in terms of angles but also distances \cite{cui_1}. Obviously, searching for the optimal codeword in a larger codebook would result in a higher pilot overhead and a lower success rate of finding the optimal codeword. \textit{This represents a qualitative rather than just quantitative difference.} Consequently, in communication systems employing ELAA, addressing the challenge of reducing pilot overhead in beam training and improving the success rate of searching for the optimal codeword, i.e., the precision of beam training, is a pressing issue.
\vspace{-0.2cm}
\subsection{State-of-the-art}
In conventional far-field communications, several beam training schemes have been proposed to diminish pilot overhead \cite{h_codebok_1, h_codebok_2, alternate_bt_1, alternate_bt_2} or to enhance the precision of beam training\cite{limin}. The authors of \cite{h_codebok_1, h_codebok_2} introuduced a classical beam training algorithm based on hierarchical codebooks to diminish the pilot overhead. Initially, the optimal wide-beam codeword is identified, followed by the determination of the optimal narrow-beam codeword within the coverage of the wide-beam codeword. Another classical beam training scheme capable of reducing pilot overhead is the alternate beam search introuduced in \cite{alternate_bt_1,alternate_bt_2}, wherein the search for the optimal transmit beam at the BS and the optimal receive beam at the user are conducted separately. Furthermore, a two-stage beam training scheme was introduced in \cite{limin}  to enhance the precision of beam training. Although the proposed scheme can satisfy the power constraints, it does not reduce the overhead. 

To meet the demands of extensive traffic and numerous connections, future 6G communication systems are anticipated to operate across multiple frequency bands, encompassing the sub-6 GHz band and the mmWave band. \cite{chengxiang_wang}.  As a matter of fact, current 5G communication systems have already exploited the full potential of the sub-6 GHz band and the mmWave band \cite{sub_mmwave_1}. The authors of \cite{hashemi} conducted extensive experiments in various indoor and outdoor settings. They demonstrated that sub-6 GHz channels and mmWave channels exhibit spatial correlation, which can be utilized to diminish the pilot overhead of mmWave beam training. Hence, to mitigate the pilot overhead required for mmWave beam training, various literature has proposed the exploitation of information from the sub-6 GHz band to aid beam training in the mmWave band. In \cite{ali_1,lizhannan}, mmWave beam training was conceptualized as a sparse signal recovery problem and the spatial parameters of the sub-6 GHz channel were extracted to assist the mmWave beam training. Moreover, the authors of \cite{ali_2} proposed to utilize the parameters of the sub-6 GHz channel to construct the mmWave channel covariance matrices, after which the mmWave channel covariance matrices can be used to assist mmWave beam training.

While the schemes proposed in \cite{hashemi, ali_1, lizhannan, ali_2} reduced the pilot overhead by exploiting knowledge of the sub-6 GHz band, they all relied on parameter estimation, which makes their performance sensitive to estimation errors. In order to alleviate this drawback, a number of deep learning-based beam training schemes were proposed\cite{mm_sub6g,simminsoo,make_sub6g,gaofeifei} to reduce or bypass parameter estimation and thus the sensitivity to estimation error is mitigated. The authors of \cite{mm_sub6g} proposed a beam training scheme utilizing fully connected neural networks (FCNN), in which FCNN is utilized to map sub-6 GHz channel state information (CSI) into optimal mmWave beam. Similarly, \cite{simminsoo} introduced an FCNN-based beam training scheme. The distinction lies in the input to the FCNN, which is the estimated power delay profile (PDP) of the sub-6 GHz channel. Furthermore, to enhance the precision of the mmWave beam training, the authors of \cite{gaofeifei} exploited not only the sub-6 GHz channels but also a few mmWave pilot signals. A novel neural network structure called FusionNet was proposed in \cite{gaofeifei} to fuse these two types of data and predict the optimal mmWave beams. Further, the authors of \cite{make_sub6g} exploited both sub-6 GHz channel and partial mmWave wide beam test information to diminish the overhead and enhance the precision of beam training, where the convolutional neural networks (CNN) were employed.

Nevertheless, the aforementioned literature solely addressed far-field channels and codebooks, making it challenging to extend them to near-field scenarios. Recognizing the distinct characteristics of near-field communication, several near-field beam training schemes have been proposed to mitigate pilot overheads \cite{luyu,self_1, self3, jiangguoli}. In \cite{luyu}, a hierarchical codebook applicable to near-field channels was introduced to decrease the pilot overheads. Nevertheless, beam training schemes relying on near-field hierarchical codebooks still necessitate substantial feedback, mirroring the requirements of far-field communication. Beyond traditional algorithms, deep learning has also been employed in beam training for near-field communication. For example, neural network based beam training schemes were proposed in \cite{self_1} and \cite{self3}. The schemes only required to test partial far-field wide beams, and then the test information was mapped by the neural network to the optimal near-field codeword. Similarly, the authors of \cite{jiangguoli} proposed to test partial near-field codewords and the test results were mapped by a CNN into optimal near-field codeword. 
\vspace{-0.2cm}

\subsection{Main Contributions}
As far as we are aware, none of the existing studies can fully harness the sub-6 GHz band to predict the optimal mmWave beam in the near-field domain. To address this gap, we introduce a beam training approach that utilizes sub-6 GHz band information to estimate the optimal mmWave beam in the near-field. In contrast to the literature which assumed that perfect sub-6 GHz channels are known \cite{mm_sub6g,make_sub6g,gaofeifei} or require parameter estimation\cite{ali_1, lizhannan, ali_2, simminsoo}, our proposed scheme only needs the uplink pilot signals received at the sub-6 GHz BS to estimate the optimal near-field mmWave beam, which is more feasible and can bypass the channel estimation. Subsequently, motivated by the success of deep learning in handling complex non-linear mapping problems, we propose utilizing neural networks to map far-field sub-6 GHz pilots to the optimal near-field mmWave codeword. \textit{Note that this mapping is not only from sub-6 GHz to mmWave but also from far field to near field, i.e., dual mapping}. Our main contribution is that we are the first to demonstrate how deep learning is suitable for this double mapping.

Furthermore, in contrast to the literature \cite{ali_1, lizhannan, ali_2, mm_sub6g,simminsoo,make_sub6g,gaofeifei,luyu,self_1, self3, jiangguoli}, we consider a user-intensive orthogonal frequency division multiplexing (OFDM) system \cite{OFDM_wangjiangzhou_1} in this paper. Leveraging user-intensive OFDM systems as a foundation, we propose a novel approach to enhance the accuracy of near-field mmWave beam training. Our method introduces a novel near-field mmWave beam estimation network, termed NMBEnet, which integrates both a CNN module and a graph neural network (GNN) module to effectively leverage the correlation among users and subcarriers. Specifically, in 6G systems characterized by high user-density, particularly in massive communication scenarios proposed by the International Telecommunication Union (ITU) \cite{ITU}, users tend to be in close proximity to each other, resulting in similar wireless propagation environments\cite{chengxiang_wang, mmwave_elaa}. Since the received signal is the product of the interaction between the transmitted signal and the wireless propagation environment\cite{jiangzhiyuan}, the sub-6 GHz uplink pilots from various users will exhibit similarities, which is called inter-user correlation. The similarity in the received signals allows not only the user’s own pilot signals to aid the BS in determining the optimal near-field mmWave beam but also enables the utilization of pilot signals from neighboring users to infer the optimal near-field mmWave codeword, resembling a form of ``diversity gain".  The GNN module included in NMBEnet can leverage correlation between users to improve beam training accuracy and can accommodate any number of users\cite{overview_gnn}. 

Moreover, since the subcarriers have similar frequencies in OFDM systems, the channels under different subcarriers would show similarities in terms of paths, delays, and angles\cite{carrier_similar_1,carrier_similar_2}. Hence the received signals at different subcarrier frequencies also exhibit similarity, which is called inter-subcarrier correlation. This correlation enables the pilot signals on each subcarrier to be used to infer user's optimal near-field mmWave beam, which can also be viewed as a sort of ``diversity gain". The CNN module included in NMBEnet can handle the correlation between subcarriers and enhance the precision of beam training. 


Our primary contributions are outlined as follows:
\begin{enumerate}
	\item We introduce a learning-based beam training scheme that utilizes only the sub-6 GHz uplink pilot signals to estimate the optimal near-field mmWave beam, which reduces the pilot overhead required for beam training in near-field mmWave communication systems and bypasses the channel estimation phase. The proposed scheme achieves accurate dual mapping from sub-6 GHz to mmWave and from far field to near field.
	\item We introduce a novel neural network termed NMBEnet designed to map sub-6 GHz uplink pilot signals to the optimal near-field mmWave beam. This neural network leverages both inter-user and inter-subcarrier correlations, effectively enhancing the precision of beam training in multiuser OFDM systems. 
	\item We present comprehensive simulation results to analyze the performance of our proposed scheme. Furthermore, we adopt the state-of-the-art simulation software called Wireless Insite (WI) in our simulation to model the multiuser OFDM communication system as realistically as possible. Our simulation results demonstrate that the proposed scheme surpasses existing beam training schemes relying on conventional neural network models and approximates the exhaustive algorithm across multiple metrics.
	
\end{enumerate}


The subsequent sections of this paper are structured as follows: Section \ref{systemmodel1} elucidates the system model adopted in this study and formulates the underlying problem. Section \ref{scheme} delineates the rationale behind and the architecture of the proposed NMBEnet. Furthermore, it presents the proposed near-field mmWave beam training scheme based on NMBEnet. Section \ref{simulation} furnishes the results of our simulations, followed by our concluding remarks in Section \ref{conclusion}.

In this paper, we employ the following notations. Bold lower case and bold upper case letters denote vectors and matrices, respectively, e.g., $\mathbf{a} $ and $\mathbf{A} $;
Scalar and a set are represented by $a $ and $\mathcal{A}  $, respectively; The $i$-th element of $\mathbf{a}$ and the $(i,j)$-th element of $\mathbf{A}$ are represented by $\left [\mathbf{a}  \right ]_{i}$ and $\left [ \mathbf{A}  \right ] _{i,j}$ , respectively; Absolute value is represented by $\left|\cdot  \right|$; Conjugate, transpose, and conjugate transpose are denoted by  $\left ( \cdot  \right )^{\ast }$, $\left ( \cdot  \right )^{\mathrm{T} }$ and $\left ( \cdot  \right )^{\mathrm{H} }$, respectively; The Gaussian distribution is denoted by $\mathcal{C} \mathcal{N}(\mu ,\sigma^{2} )$ where $\mu$ is mean and $\sigma^{2}$ is variance. The Frobenius norm is denoted by $\left \|  \cdot \right \| _{\mathrm{F} } $.

%
%


\section{System Model}\label{systemmodel1}
\begin{figure}[t]
	\centering
	\includegraphics[width=2.5in]{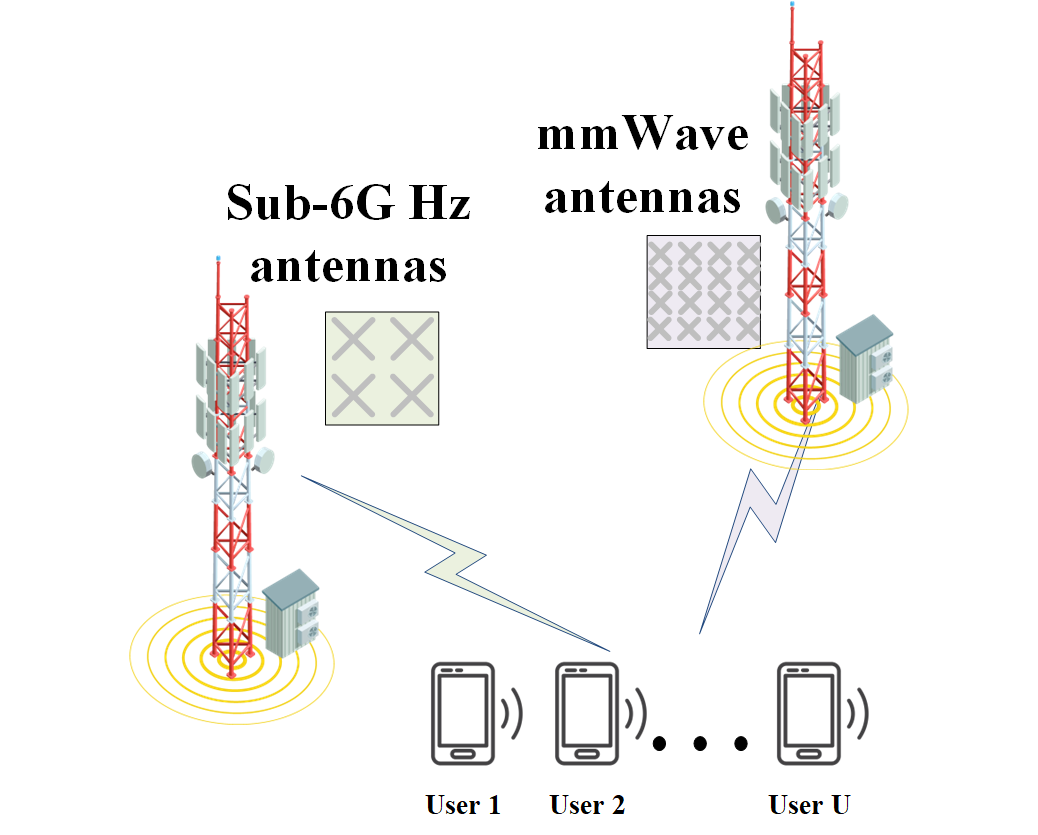}
	\vspace{-0.2cm}
	\caption{\fontsize{11pt}{\baselineskip}\selectfont Illustration of the adopted system model.}
	\label{system_model}
	\vspace{-0.6cm}
\end{figure}

Consider a multi-user OFDM communication system consisting of a sub-6 GHz BS, a mmWave BS and $U$ users, which is shown in Fig. \ref{system_model}. Note that each variable for the sub-6 GHz band are added an upper line to distinguish them from those for the mmWave band. The mmWave BS is deployed with ELAA containing $M$-antenna uniform linear array \footnote{Our proposed scheme can be easily adaptable to systems employing uniform planar arrays (UPA). For UPAs, the scheme needs substituting the 2-dimensional near-field codebook with a 3-dimensional counterpart and employing separate networks to determine the azimuth and elevation angle indices.}
and $N$ radio frequency (RF) chains, where $U\le N\le M$ is satisfied. To save transmission power, it is commonly assumed that redundant RF chains are switched off, so that the number of RF chains is equal to the number of users, i.e., $N=U$\cite{sun,hybrid_precoding_AA_1}. Since the number of RF chains is limited, mmWave BS is assumed to employ hybrid precoders, which consist of an analog precoder and a digital precoder. The sub-6 GHz BS is equipped with $\overline{M}$ antennas. Since $\overline{M}$ is generally small, the sub-6 GHz BS is assumed to be fully digital, where each antenna is connected to an independent RF chain. Furthermore, each user is equipped with a sub-6 GHz antenna and a mmWave antenna.

	\vspace{-0.3cm}
\subsection{mmWave Signal Model}
For mmWave links, we consider a time division duplex (TDD)-based multiuser OFDM signal model, where $K$ OFDM subcarriers are used for data transmissions to tackle frequency selective fading. Let us denote the signal vector transmitted on the $k$-th subcarrier by $\mathbf{s}\left[ k \right] =\left[ s_1\left[ k \right] ,s_2\left[ k \right] ,\dots ,s_U\left[ k \right] \right] ^{\mathrm{T}}$, where $s_u\left[ k \right]$ denotes the symbol transmitted to user $u$. Before transmission, the signal vectors on each subcarrier are first precoded by the digital precoder, after which a cyclic prefix of length $L$ is appended and the signals are transformed to the time domain by using $K$-point inverse fast Fourier transforms (IFFTs). Finally, the transformed signals are precoded by the analog precoder and then transmitted by the antenna array. Note that the analog precoding follows after the IFFT, hence the analog precoder is the same for signals on all subcarriers. 

After the signals arrive at the users, the received signals are transformed to the frequency domain by FFT and the cyclic prefixes are removed. Consider a block-fading channel model, the signal received by user $u$ on the $k$-th subcarrier can be represented as
\begin{equation}\label{mmwave_dl_r}
	\setlength\abovedisplayskip{3pt}
	\setlength\belowdisplayskip{3pt}
	\begin{aligned}
     y_{u}^{\mathrm{dl}}\left[ k \right]& =\mathbf{h}_{u}^{\mathrm{dl}}\left[ k \right] \mathbf{F}_{\mathrm{RF}}\mathbf{F}_{\mathrm{BB}}\left[ k \right] \mathbf{s}\left[ k \right] +n_{u}^{\mathrm{dl}}\left[ k \right] \\
     &=\mathbf{h}_{u}^{\mathrm{dl}}\left[ k \right] \mathbf{F}_{\mathrm{RF}}\sum_{u=1}^U{\mathbf{f}_{u}^{\mathrm{BB}}\left[ k \right] s_u\left[ k \right]}+n_{u}^{\mathrm{dl}}\left[ k \right],
	\end{aligned}
\end{equation}
where $\mathbf{h}_{u}^{\mathrm{dl}}\left[ k \right] \in \mathbb{C} ^{1\times M}$ denotes the downlink channel of the $u$-th user on the $k$-th subcarrier, $\mathbf{F}_{\mathrm{RF}}=\left[ \mathbf{f}_{1}^{\mathrm{RF}},\mathbf{f}_{2}^{\mathrm{RF}},\dots ,\mathbf{f}_{U}^{\mathrm{RF}} \right] \in \mathbb{C} ^{M\times N}$ and $\mathbf{F}_{\mathrm{BB}}\left[ k \right] =\left[ \mathbf{f}_{1}^{\mathrm{BB}}\left[ k \right] ,\mathbf{f}_{2}^{\mathrm{BB}}\left[ k \right] ,\dots ,\mathbf{f}_{U}^{\mathrm{BB}}\left[ k \right] \right] \in \mathbb{C} ^{N\times N} $ denote the analog precoder and the digital precoder, respectively. Since the analog precoding is implemented by the phase shifters, the elements of $\mathbf{F}_{\mathrm{RF}}$ satisfy the constraint of constant modulus, i.e. $\left|\left [ \mathbf{F}_{\textrm{RF}} \right ] _{m,n}\right|=\frac{1}{\sqrt{M}}$. To satisfy the power constraints, $\left\| \mathbf{F}_{\mathrm{RF}}\mathbf{f}_{u}^{\mathrm{BB}}\left [ k \right ]  \right\| _{\mathrm{F}}^{2}=1,u=1,2,...,U.$ and $\mathrm{E}\left[ \mathbf{s}\left[ k \right] \mathbf{s}\left[ k \right] ^{\mathrm{H}} \right] =\frac{P_{\mathrm{dl}}}{U}\mathbf{I}_U $ are satisfied, where $P_{\mathrm{dl}}$ denotes the downlink transmission power assigned to each subcarrier. Moreover, $n_{u}^{\mathrm{dl}}\left[ k \right] \sim \mathcal{C} \mathcal{N} \left( 0,\sigma _{\mathrm{dl}}^{2}\right) $ denotes additive white Gaussian noise, where $\sigma _{\mathrm{dl}}^{2}$ denotes the downlink noise power.

Based on the signal model given in (\ref{mmwave_dl_r}), the spectral efficiency of the $u$-th user at the $k$-th subcarrier is expressed as
\begin{equation}\label{sum_rate}
	\setlength\abovedisplayskip{3pt}
	\setlength\belowdisplayskip{3pt}
	\begin{split}
		R_u\left[ k \right] =\log _2\left( 1+\frac{\frac{P_{\mathrm{dl}}}{U}\left| \mathbf{h}_{u}^{\mathrm{dl}}\left[ k \right] \mathbf{F}_{\mathrm{RF}}\mathbf{f}_{u}^{\mathrm{BB}}\left[ k \right] \right|^2}{\frac{P_{\mathrm{dl}}}{U}\sum\nolimits_{i\ne u}^{}{\left| \mathbf{h}_{i}^{\mathrm{dl}}\left[ k \right] \mathbf{F}_{\mathrm{RF}}\mathbf{f}_{i}^{\mathrm{BB}}\left[ k \right] \right|^2+\sigma _{\mathrm{dl}}^{2}}} \right).
	\end{split}
\end{equation}

\vspace{-0.5cm}
\subsection{mmWave Channel Model}
Due to the deployment of ELAA, the boundary between the near field and the far field, i.e., the Rayleigh distance, is expanded. Therefore, near-field communication based on the spherical wave assumption has to be taken into account, because users are more likely to be situated within the near-field domain.
To characterize the near-field mmWave channel, we adopt the near-field wideband geometric channel model \cite{wangmingjin, linyuxing}. We derive the uplink channel first, after which the downlink channel can be easily obtained by transposing the uplink channel in TDD systems. We use $\alpha \left[ i \right] $ to represent the discrete-time baseband transmitted signal having sampling period $T_s=\frac{1}{f_s}$. Then, the continuous time baseband transmitted signals can be obtained by
 \begin{equation}\label{baseband_signal}
 	\setlength\abovedisplayskip{3pt}
 	\setlength\belowdisplayskip{3pt}
 	\begin{split}
 	x\left( t \right) =\sum_{i=-\infty}^{+\infty}{\alpha \left[ i \right] g\left( t-iT_s \right)},
   \end{split}
 \end{equation}
where $g\left( t \right) $ is the pulse shaping function. After modulation, the transmitted passband signal can be represented as
 \begin{equation}\label{passband_signal}
	\setlength\abovedisplayskip{3pt}
	\setlength\belowdisplayskip{3pt}
	\begin{split}
	\widetilde{x}\left( t \right) =\mathfrak{R} \left\{ x\left( t \right) e^{j2\pi f_ct} \right\},
	\end{split}
\end{equation}
where $f_c$ denotes the centre carrier frequency. The number of paths between the user and the mmWave BS is denoted by $L$ and the complex channel gain of each path is denoted by $\beta _l$. Let $\tau _l$ denote the propagation delay from the user to the centre antenna of the mmWave BS. The extra propagation delay of the $m$-th antenna compared to the central antenna of the $l$-th path is denoted by $\varDelta _{l,m}$. From \cite{cui_1}, $\varDelta _{l,m}$ can be obtained by
 \begin{equation}\label{extra_delay}
	\setlength\abovedisplayskip{3pt}
	\setlength\belowdisplayskip{3pt}
	\begin{split}
	\varDelta _{l,m}=\frac{r_l-r_{l}^{\left( m \right)}}{c}=\frac{r_l-\sqrt{r_{l}^{2}-\sigma _{m}^{2}d^2-2r_l\theta _l\sigma _md}}{f_c\lambda _c},
	\end{split}
\end{equation}
where $r_l $ denotes the distance from the centre antenna to the user or the last scatterer on the $l$-th path, $r_{l}^{\left( m \right)}$ denotes the distance from the $m$-th antenna to the user or the last scatterer on the $l$-th path. $\sigma _m=\frac{2m-M+1}{2},m=0,1,\dots ,M-1$ denotes the index of the antenna. $\theta _l$ denotes the sine of the arrival angle of the central antenna on the $l$-th path. $c$, $\lambda _c$ and $d$ denote the speed of light, wavelength and antenna spacing, respectively, where $d$ is assumed to be $\lambda _c/2$. Subsequently, the baseband received signal at the $m$-th antenna can be represented as
\begin{equation}\label{time_received_signal}
	\setlength\abovedisplayskip{3pt}
	\setlength\belowdisplayskip{3pt}
	\begin{split}
	y_m\left( t \right) &=\sum_{l=1}^L{\beta _l x\left( t-\tau _l-\varDelta _{l,m} \right)}e^{-j2\pi f_c\left( \tau _l+\varDelta _{l,m} \right)}
	\\
	&=\left( \sum_{l=1}^L{\widetilde{\beta }_le^{-j2\pi f_c\varDelta _{l,m}}\delta \left( t-\tau _l-\varDelta _{l,m} \right)} \right) \ast x\left( t \right) ,
	\end{split}
\end{equation}
where $\delta \left( t \right) $ is the impulse signal and $\widetilde{\beta }_l=\beta _le^{-j2\pi f_c\tau _l}$ denotes the equivalent complex channel gain. Note that in systems with massive antenna arrays, it is commonly assumed that $x\left( t-\tau _l-\varDelta _{l,m} \right) \approx x\left( t-\tau _l \right) $ because that $\varDelta _{l,m}\ll T_{\mathrm{s}}$ is always stisfied for each $m=0,1,\dots ,M-1$. However, this approximation will no longer hold in systems containing ELAA because of the large value of $M$. After performing the Fourier transform on $y_m\left( t \right)$, the frequency domain signal can be obtained by
\begin{equation}\label{frequency_received_signal}
	\setlength\abovedisplayskip{3pt}
	\setlength\belowdisplayskip{3pt}
	\begin{split}
	y_m\left( f \right) &=\int_{-\infty}^{+\infty}{y_m\left( t \right)}e^{-j2\pi ft}dt
	\\
	&=\left( \sum_{l=1}^L{\widetilde{\beta }_le^{-j2\pi f_c\varDelta _{l,m}}e^{-j2\pi f\left( \tau _l+\varDelta _{l,m} \right)}} \right) x\left( f \right) 
\\
	&=\left( \sum_{l=1}^L{\widetilde{\beta }_le^{-j2\pi f_c\tau _l}e^{-j2\pi \left( 1+\frac{f}{f_c} \right) \phi _{l,m}}} \right) x\left( f \right) ,
	\end{split}
\end{equation}
where $\phi _{l,m}=f_c\varDelta _{l,m}=\frac{r_l-r_{l}^{\left( m \right)}}{\lambda _c}$. Then the received signal vector at $M$ antennas can be represented as
\begin{equation}\label{received_signal_vector}
	\setlength\abovedisplayskip{3pt}
	\setlength\belowdisplayskip{3pt}
	\begin{split}
	\mathbf{y}\left( f \right) &=\left[ y_1\left( f \right) ,y_2\left( f \right) ,\dots ,y_M\left( f \right) \right] ^{\mathrm{T}}
	\\
&	=\mathbf{h}^{\mathrm{ul}}\left( f \right) x\left( f \right) ,
	\end{split}
\end{equation}
and the channel model can be expressed as 
\begin{equation}\label{channel_model}
	\setlength\abovedisplayskip{3pt}
	\setlength\belowdisplayskip{3pt}
	\begin{split}
		\mathbf{h}^{\mathrm{ul}}\left( f \right) =\sum_{l=1}^L{\widetilde{\beta }_le^{-j2\pi f_c\tau _l}}\mathbf{b}\left( \theta _l,r_l \right),
	\end{split}
\end{equation}
where 
\begin{equation}\label{near_steering_vector}
	\begin{aligned}
		\mathbf{b}\left( \theta _l,r_l \right) =[ e^{-j2\pi \left( 1+\frac{f}{f_c} \right) \phi _{l,1}},e^{-j2\pi \left( 1+\frac{f}{f_c} \right) \phi _{l,2}},	\dots \\
		,e^{-j2\pi \left( 1+\frac{f}{f_c} \right) \phi _{l,M}} ] ,
	\end{aligned}
\end{equation}
According to (\ref{channel_model}) and (\ref{near_steering_vector}), the downlink channel on the $k$-th subcarrier can be represented as
\begin{equation}\label{downlink_channel}
	\setlength\abovedisplayskip{3pt}
	\setlength\belowdisplayskip{3pt}
	\begin{split}
	\mathbf{h}^{\mathrm{dl}}\left[ k \right] =\left( \mathbf{h}^{\mathrm{ul}}\left( f_k \right) \right) ^{\mathrm{T}},
	\end{split}
\end{equation}
where $f_k$ denotes the $k$-th subcarrier frequency.
\vspace{-0.3cm}
\subsection{sub-6 GHz Signal Model}
For sub-6 GHz links, we also consider a TDD-based multiuser OFDM signal model, where $\overline{K}$ OFDM subcarriers are used on the uplink. Furthermore, $U$ users are assumed to transmit mutually time-orthogonal uplink pilot signals to the sub-6 GHz BS. Since the sub-6 GHz BS is fully digital, the received pilot signals of the $u$-th user on the $\overline{k}$-th subcarrier can be represented as
\begin{equation}\label{sub6g_ul_r}
	\setlength\abovedisplayskip{3pt}
	\setlength\belowdisplayskip{3pt}
	\begin{split}
		\overline{\mathbf{y}}_u[\overline{k}]=\overline{\mathbf{h}}_{u}^{\mathrm{ul}}[\overline{k}]z_u[\overline{k}]+\overline{\mathbf{n}}_{u}^{\mathrm{ul}}[\overline{k}],
	\end{split}
\end{equation}
where $z_u[ \overline{k}] \in \mathbb{C} $ and $\overline{\mathbf{h}}_{u}^{\mathrm{ul}}[ \overline{k} ] \in \mathbb{C} ^{\overline{M}\times 1}$ denote the transmitted uplink pilot symbol and uplink channel, respectively. $z_u[ \overline{k}]  $ is assumed to satisfy the power constraint $\left | z_u[ \overline{k}] \right | ^{2} =\overline{P} _{\mathrm{ul} } $, where $\overline{P} _{\mathrm{ul} }$ denotes the uplink transmission power on each subcarrier. Furthermore, $\overline{\mathbf{n}}_{u}^{\mathrm{ul}}[\overline{k}] \sim \mathcal{C} \mathcal{N} \left( \mathbf{0},\overline{\sigma }_{\mathrm{ul}}^{2}\mathbf{I} \right) $ denotes additive white Gaussian noise, where $\overline{\sigma }_{\mathrm{ul}}^{2}$ denotes the uplink noise power.

\vspace{-0.3cm}
\subsection{sub-6 GHz Channel Model}
For sub-6 GHz systems, users are more likely to locate in the far-field domain of sub-6 GHz systems due to the smaller size of the antenna array and the smaller Rayleigh distance. Consider a wideband geometric channel model for the far field, which can be represented as
\begin{equation}\label{sub6g_ul_channel}
	\setlength\abovedisplayskip{3pt}
	\setlength\belowdisplayskip{3pt}
	\begin{split}
		\overline{\mathbf{h}}^{\mathrm{ul}}\left( f \right) =\sum_{\overline{l}=1}^{\overline{L}}{\overline{\beta }_le^{-j2\pi \overline{f}_c\overline{\tau }_{\overline{l}}}}\mathbf{a}\left( \overline{\theta }_{\overline{l}},f \right) ,
	\end{split}
\end{equation}
where $\overline{L}$ denotes the number of paths. $\overline{\beta }_l$, $\overline{\tau }_{\overline{l}}$ and $\overline{\theta }_{\overline{l}}$ denote the complex channel gain, free-space propagation delay and the sine of arrival angle on the $\overline{l}$-th path, respectively. Furthermore, the far-field steering vector $\mathbf{a}\left( \overline{\theta }_{\overline{l}},f \right) $ is represented as \cite{linyuxing, wangmingjin}
\begin{equation}\label{far_steering_vector}
	\setlength\abovedisplayskip{3pt}
	\setlength\belowdisplayskip{3pt}
	\begin{split}
		\mathbf{a}\left( \overline{\theta }_{\overline{l}} \right) =\left[ 1,e^{-j\frac{2\pi \overline{d}}{\overline{\lambda _c}}\overline{\theta }_{\overline{l}}},\dots ,e^{-j\left( \overline{M}-1 \right) \frac{2\pi \overline{d}}{\overline{\lambda _c}}\overline{\theta }_{\overline{l}}} \right] .
	\end{split}
\end{equation}

Based on the above channel model, the sub-6 GHz uplink channel on the $\overline{k}$-th subcarrier can be written as
\begin{equation}\label{uplink_channel}
	\setlength\abovedisplayskip{3pt}
	\setlength\belowdisplayskip{3pt}
	\begin{split}
	\overline{\mathbf{h}}_{}^{\mathrm{ul}}[ \overline{k} ] =\overline{\mathbf{h}}^{\mathrm{ul}}\left( \overline{f}_{\overline{k}} \right).
	\end{split}
\end{equation}

	\vspace{-0.3cm}
\subsection{Problem Formulation}
Our main objective is to design the analog precoder and digital precoder based on the sub-6 GHz uplink pilot signals to maximize the sum of the spectral efficiency of the mmWave system, i.e. $ {\textstyle \sum_{u}^{}}  {\textstyle \sum_{k}^{}}R_u\left[ k \right]$.

For the analog precoder, since the hardware constraint of the RF that the phase shifters can only use quantized angles, the column vectors of the analog precoder have to be selected from a predefined finite-size codebook\cite{hybrid_precoding_AA_1, sun}. The discrete Fourier transform (DFT) codebook has been widely used in far-field communication to design the analog precoder. However, the DFT codebook is not applicable to near-field communications, as the assumption of planar-waves no longer holds in the near-field domain. Therefore, we adopt the near-field codebook based on the spherical wave assumption proposed in \cite{cui_1}, which incorporates not only the search for angle but also for distance. In particular, the space is uniformly divided into $M$ angles in terms of direction by the near-field codebook, while the space is divided into $S$ distance rings in terms of distance. Then, the near-field codebook is given by
\begin{equation}\label{near_codebook}
	\setlength\abovedisplayskip{3pt}
	\setlength\belowdisplayskip{3pt}
	\begin{split}
	\mathcal{N} =\left\{ \mathbf{b}\left( \psi  _1,r_{1,1} \right) ,\dots ,\mathbf{b}\left( \psi _M,r_{1,M} \right) ,\dots ,\mathbf{b}\left( \psi_M,r_{S,M} \right) \right\} .
\end{split}
\end{equation}
where $\psi _m$ denotes the $m$-th sampling angle and $r_{s,m}$ denotes the sampling distance from the BS to the $s$-th distance ring under the angle $\psi _m$.

Then the problem of maximizing the sum spectral efficiency is given by
\begin{equation}\label{problem_1}
	\setlength\abovedisplayskip{3pt}
	\setlength\belowdisplayskip{3pt}
	\begin{split}
		&\underset{\mathbf{F}_{\mathrm{RF}},\mathbf{F}_{\mathrm{BB}}\left[ k \right]}{\max}\sum_{u=1}^U{\sum_{k=1}^K{R_u\left[ k \right]}}
		\\
		&\mathrm{s}.\mathrm{t}. \ \left[ \mathbf{F}_{\mathrm{RF}} \right] _{:,u}=\mathbf{f}_{u}^{\mathrm{RF}}\in \mathcal{N} ,    u=1,2,\dots U
		\\
		&\,\,    \left\| \mathbf{F}_{\mathrm{RF}}\mathbf{f}_{u}^{\mathrm{BB}}\left[ k \right] \right\| _{\mathrm{F}}^{2}=1,    u=1,2,\dots ,U\,\, k=1,2,\dots K.
	\end{split}
\end{equation}

Note that solving the optimization problem given in (\ref{problem_1}) is difficult because it is non-convex. Depending on the existing contributions\cite{sun,hybrid_precoding_AA_1,hybrid_precoding_AA_2,hybrid_precoding_3}, a common approach to solving Problem (\ref{problem_1}) is to design the analog precoder and digital precoder separately. Specifically, we first need to select the optimal codewords for each user to form the analog precoder with a fixed digital precoder, after which the digital precoder is determined. When determining the analog precoder, the problem of selecting the optimal codeword for each user can be written as
\begin{equation}\label{problem_2}
	\setlength\abovedisplayskip{3pt}
	\setlength\belowdisplayskip{3pt}
	\begin{split}
	&\underset{\mathbf{f}_{u}^{\mathrm{RF}}}{\max}\sum_{k=1}^K{\left| \mathbf{h}_{u}^{\mathrm{dl}}\left[ k \right] \mathbf{f}_{u}^{\mathrm{RF}} \right|},    u=1,2\dots U
	\\
	&\mathrm{s}.\mathrm{t}. \ \mathbf{f}_{u}^{\mathrm{RF}}\in \mathcal{N} ,    u=1,2,\dots U,
	\end{split}
\end{equation}
where $g_u=\sum_{k=1}^K{\left| \mathbf{h}_{u}^{\mathrm{dl}}\left[ k \right] \mathbf{f}_{u}^{\mathrm{RF}} \right|} $ represents the sum beamforming gain from $\mathbf{f}_{u}^{\mathrm{RF}}$ for the $u$-th user on all subcarriers. It can be seen that Problem (\ref{problem_2}) is a typical beam training problem. Since there are already established schemes for the design of digital precoders\cite{sun, hybrid_precoding_AA_1,hybrid_precoding_AA_2}, e.g., using zero-forcing (ZF) or minimum mean-squared error (MMSE) algorithms, hence we mainly focus on how to design a beam training scheme to solve Problem (\ref{problem_2}) in this paper.

Since near-field codebooks contain not only the search in angle but also in distance, the number of codewords in the near-field codebook increases dramatically. Searching for the optimal codeword within an extensive codebook inherently leads to increased pilot overhead and diminished likelihood of finding the optimal codeword. This distinction is not only quantitative but also qualitative in nature. Consequently, in communication systems employing ELAA, the imperative challenge lies in minimizing pilot overhead during beam training while concurrently enhancing the success rate of finding the optimal codeword, i.e., the accuracy of beam training.

\section{NMBEnet-based beam training with dual mapping}\label{scheme}
\subsection{Motivation}
\begin{figure}[t]
	\centering
	\includegraphics[width=3in]{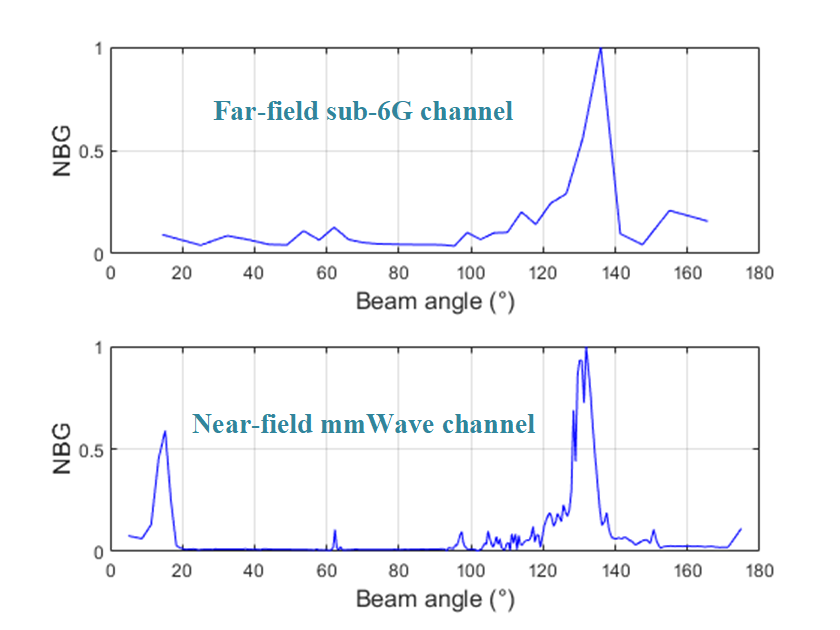}
	\vspace{-0.2cm}
	\caption{\fontsize{11pt}{\baselineskip}\selectfont Similarity between beam patterns of far-field sub-6 GHz channel and near-field mmWave channel.}
	\label{sub_mm}
	\vspace{-0.6cm}
\end{figure}

Since the existing literature has demonstrated the spatial correlation between sub-6 GHz channels and mmWave channels\cite{sub_mm_correlation_1,sub_mm_correlation_2}, the information of sub-6 GHz bands can be used to aid beam training of the mmWave bands\cite{mm_sub6g}. Moreover, according to the definition of the Rayleigh distance \cite{def_rayleigh}, the near-field spherical wave model can be approximated to the far-field planar wave model when the maximum phase discrepancy among all antennas is less than $\pi /8$. This approximation suggests that there are also correlations between the near-field channel and the far-field channel, enabling the utilization of far-field information for near-field beam training. In addition, we use the latest simulation software called WI to obtain the far-field sub-6 GHz channel and the near-field mmWave channel to validate this correlation\cite{WI}. For a more intuitive view of the channel's characteristics, the channel is left-multiplied by the DFT matrix and normalized to obtain the beam pattern of the channel which is shown in Fig. \ref{sub_mm} where the horizontal coordinate is the beam angle and the vertical coordinate is the normalized beamforming gain (NBG) for each beam. From Fig. \ref{sub_mm}, we can readily observe the correlation between the far-field sub-6 GHz channel and the near-field mmWave channel.

Motivated by the above, we adopt the information of far-field sub-6 GHz to assist in the search for optimal near-field mmWave codewords, which can significantly reduce the pilot overheads. Not only that, in order to reduce the complexity, we bypass the channel estimation and directly use the far-field sub-6 GHz pilot signals to assist the far-field mmWave beam training, because the uplink pilot signals are shown to be sufficient for solving some optimisation problems for downlink\cite{kareem}. This beam training model can be expressed mathematically as 
\begin{equation}\label{beamtraining_1}
	\setlength\abovedisplayskip{3pt}
	\setlength\belowdisplayskip{3pt}
	\begin{split}
	\left\{ \mathbf{b} _{u}^{\star }  \right \} =f_{m} (\overline{\mathbf{y}}_u[\overline{k}]),   u=1,2, \dots,U, 
	\end{split}
\end{equation}
where $\mathbf{b} _{u}^{\star }$ denotes the index of the optimal near-field mmWave codeword for the $u$-th user and $f_{m}\left ( \cdot  \right ) $ denotes the mapping function. However, this beam training model, which relies only on a single user's pilot signals at a single carrier, was shown to have poor accuracy\cite{make_sub6g}.

In future 6G communication systems that are expected to deploy ELAA,  the utilization of OFDM technology will persist and user-intensive scenarios will be common\cite{chengxiang_wang, ITU}. Consequently, it becomes imperative to address user-intensive OFDM systems. Leveraging inter-user and inter-subcarrier correlations presents an opportunity to enhance beam training accuracy. 
\begin{figure*}
	\vspace{-0.1cm}
	\begin{minipage}[t]{0.45\linewidth}
	\centering
	\includegraphics[width=3in]{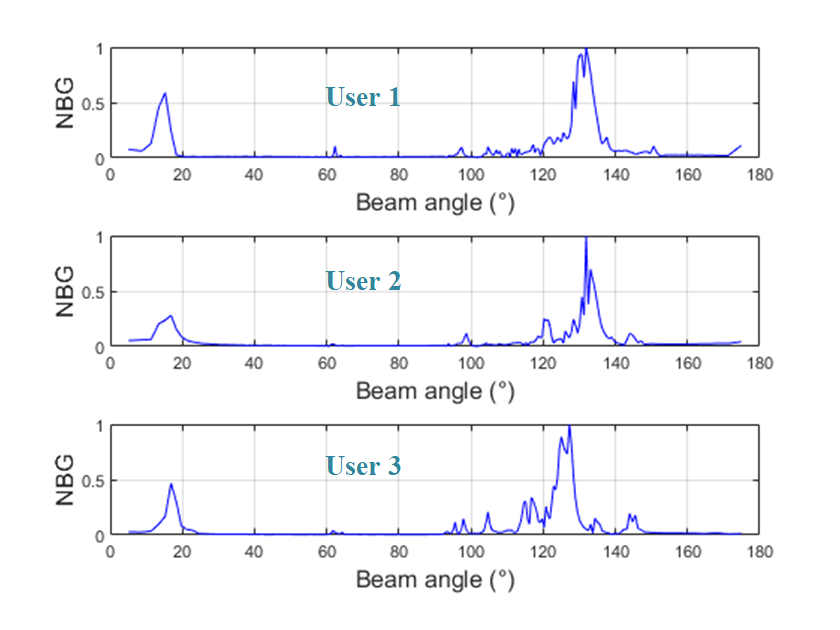}
	\vspace{-0.2cm}
	\caption{\fontsize{11pt}{\baselineskip}\selectfont Similarity between the beam patterns of different users' channels.}
	\label{user_user}\vspace{-0.2cm}
	\end{minipage}%
	\hfill
	\begin{minipage}[t]{0.45\linewidth}
			\centering
		\includegraphics[width=3in]{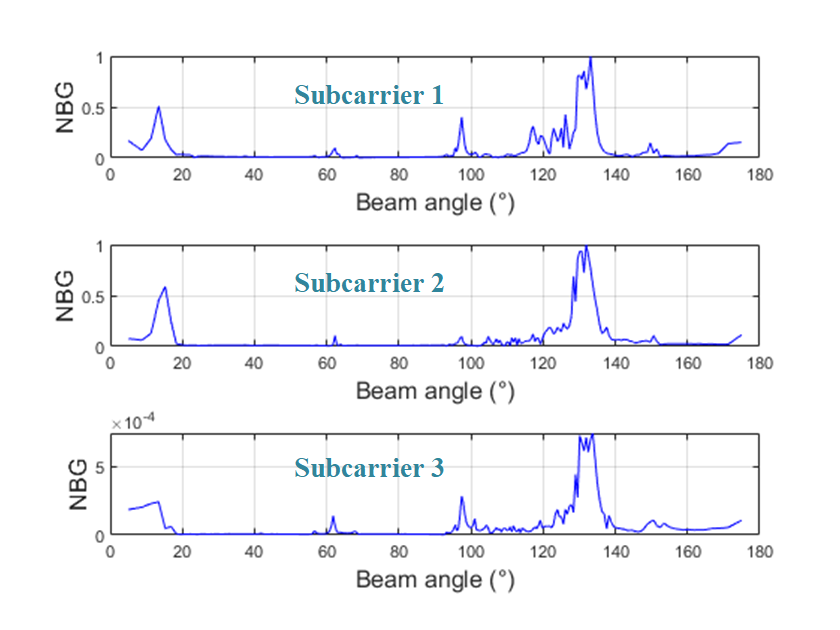}
		\vspace{-0.2cm}
		\caption{\fontsize{11pt}{\baselineskip}\selectfont Similarity between beam patterns of channels with different subcarriers.  }
		\label{fc_fc}\vspace{-0.2cm}
	\end{minipage}%
	\hfill
	\vspace{-0.2cm}
\end{figure*}
Specifically, in 6G systems, particularly in high user-density environments such as those outlined in ITU's proposed massive communication scenarios, users are often in close proximity, leading to comparable wireless propagation environments thus similar channels. We perform simulations using WI to obtain the channels of different users in a user-intensive scenario and the beam patterns are shown in Fig. \ref{user_user}, through which similarity of the channels can be observed. As the received signal is influenced by the interaction between the transmitted signal and the wireless propagation environment, uplink pilot signals from various users exhibit inter-user correlation. Leveraging this similarity in received signals, both the user's own pilot signals and those of neighbouring users can aid in determining the optimal near-field mmWave beam, akin to a form of ``diversity gain".

Furthermore, in OFDM systems, the similarity in frequencies among subcarriers results in analogous characteristics across channels, including path, delay, and angle\cite{carrier_similar_1,carrier_similar_2}. Through the beam patterns of the channels at different subcarrier frequencies in Fig. \ref{fc_fc}, we can observe this inter-channel similarity. Consequently, received signals across different subcarrier frequencies would also demonstrate inter-subcarrier correlation. This correlation facilitates the utilization of pilot signals on each subcarrier to deduce the optimal near-field mmWave beam for users, which is also akin to a form of ``diversity gain".

In order to take advantage of inter-user and inter-subcarrier correlations, the beam training scheme is rewritten as
\begin{equation}\label{beamtraining_2}
	\setlength\abovedisplayskip{3pt}
	\setlength\belowdisplayskip{3pt}
	\begin{split}
	&	\left\{ \mathbf{b} _{1}^{\star },\mathbf{b} _{2}^{\star },\dots ,\mathbf{b} _{U}^{\star }  \right \} \\&=f_{m} \left ( \left \{ \overline{\mathbf{y}}_1[\overline{k}] \right \}_{\overline{k}=1}^{\overline{K}} ,\left \{ \overline{\mathbf{y}}_2[\overline{k}] \right \}_{\overline{k}=1}^{\overline{K}},\dots ,\left \{ \overline{\mathbf{y}}_U[\overline{k}] \right \}_{\overline{k}=1}^{\overline{K}}\right )   ,
	\end{split}
\end{equation}
which exploits the pilot signals of all users on all subcarriers simultaneously. Here it is important to note the difference between (\ref{beamtraining_1}) and (\ref{beamtraining_2}): in (\ref{beamtraining_1}) the beam training for each user is separate and only a single carrier is considered, which is the common idea adopted in the existing literature\cite{mm_sub6g,simminsoo,make_sub6g,gaofeifei}, while in (\ref{beamtraining_2}) the beam training for all the users is performed simultaneously and the pilot signals on all the subcarriers are employed. However, the mapping function in the beam training model (\ref{beamtraining_2}) is difficult to implement because not only is the relationship between the far-field sub-6 GHz signal and the near-field mmWave beam highly non-linear, but also the two correlations are difficult to extract and exploit.

In order to implement the mapping function in (\ref{beamtraining_2}), we decide to adopt the deep learning approach and conceive a neural network called NMBEnet, whose package mainly contains a CNN module and a GNN module, for the following main reasons. Firstly, the relationship between $b_{u}^{\star }$ and $\left \{ \overline{\mathbf{y}}_u[\overline{k}] \right \}_{\overline{k}=1}^{\overline{K}}$ is highly non-linear, which makes traditional estimation methods difficult to work or impose unacceptable computational complexity\cite{make1,make_sub6g}. For this reason, deep learning with its strong ability to learn non-linear relationships is employed to realize this tricky mapping task\cite{mm_sub6g}. However, traditional FCNNS still struggles to achieve the mapping in (\ref{beamtraining_2}) because they cannot exploit the two correlations. 

Due to the excellent performance of CNNs on the image classification task\cite{CNN}, we decide to use CNNs to do the initial feature extraction on the sub-6 GHz pilot signal, in which the inter-subcarrier correlation is explored. Specifically, given the finite number of codewords, beam training can be conceptualized as a classification task, with each codeword representing a category\cite{make1}. Moreover, since the sub-6 GHz pilot signals $\left \{ \overline{\mathbf{y}}_u[\overline{k}] \right \}_{\overline{k}=1}^{\overline{K}}$ share many of the same features with the image, the sub-6 GHz pilot signals can be regarded as an image with $\overline{K}$ layers, which is adopted by much of the existing literature\cite{pilot_image}. The convolution kernel in the CNN can exploit the correlation between image layers and extract features from multiple layers of the image simultaneously\cite{CNN}. Hence, the CNN can also exploit the correlation of sub-6 GHz pilot signals under different frequencies, i.e., the inter-subcarrier correlations.

Furthermore, we also consider GNN because it can leverage the correlation between users. Specifically, unlike FCNNs, GNNs have a unique structure of ``combination" and ``aggregation", which makes it feasible to take into account the features of the other users when updating  user's own features\cite{overview_gnn}. Therefore, the GNN can search for optimal codewords based on the features extracted by the CNN and leverage the correlation between users in the process.

\vspace{-0.3cm}

\subsection{ Architecture of the NMBEnet}
\begin{figure*}[t]
	\centering
	\includegraphics[width=7.2in]{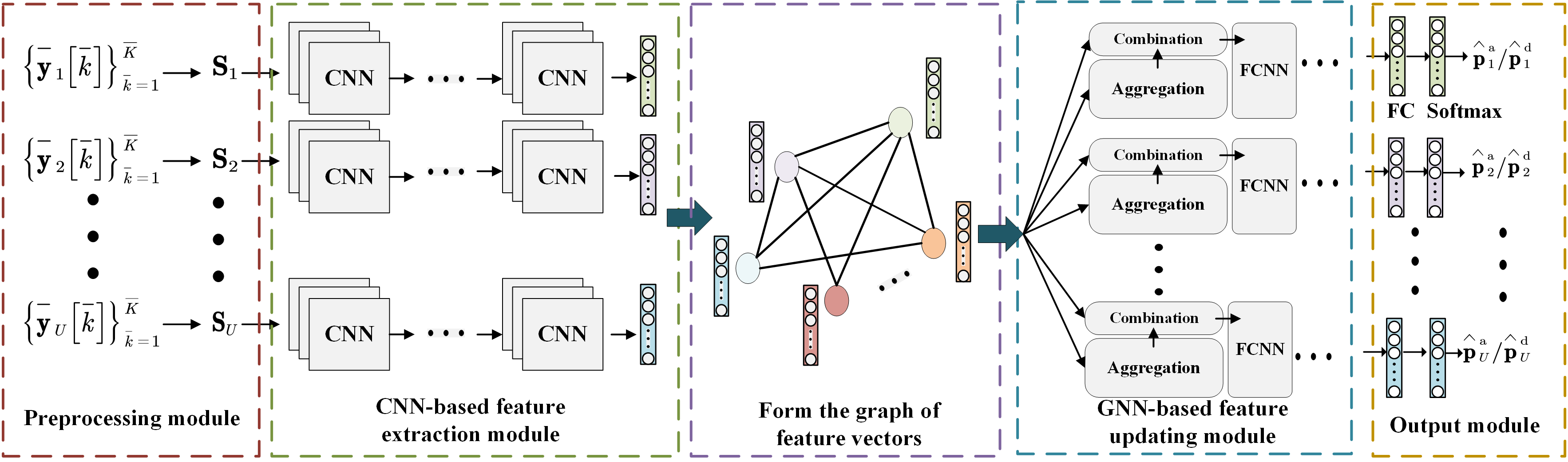}
	\vspace{-0.2cm}
	\caption{\fontsize{11pt}{\baselineskip}\selectfont Overall architecture of the poposed NMBEnet.}
	\label{NMBEnet}
	\vspace{-0.6cm}
\end{figure*}

The complete architecture of our proposed NMBEnet is shown in Fig. \ref{NMBEnet}, which consists of a preprocessing module, a CNN-based feature extraction module, a GNN-based feature updating module, and an output module.

$1)$ $\textit{Preprocessing Module}$:
Since neural networks can only handle real values, we first have to convert the uplink sub-6 GHz pilot signals in complex values to real values. Furthermore, in order to better exploit the advantages of CNNs for image processing and to fully explore the correlation between subcarriers, we treat the pilot signals as an ``image" and take the pilot signals under each subcarrier as a layer of the input ``image". This preprocessing step is given by
\begin{equation}\label{complex_to_real}
	\setlength\abovedisplayskip{3pt}
	\setlength\belowdisplayskip{3pt}
	\begin{split}
		\mathbf{S}_{u}  \left [ \overline{k} \right ]= \left [ \mathfrak{R} \left \{ \overline{\mathbf{y}}_u[\overline{k}] \right \},\mathfrak{I} \left \{ \overline{\mathbf{y}}_u[\overline{k}] \right \} \right ]    ,
	\end{split}
\end{equation}
where $\mathbf{S}_{u}\in \mathbb{C} ^{\overline{K}\times 2\times \overline{M}} $ is the input of the $u$-th user and $	\mathbf{S}_{u}  \left [ \overline{k} \right ]$ is the $\overline{K}$-th layer of the input. $\mathfrak{R} \left ( \cdot  \right ) $ and $\mathfrak{I} \left ( \cdot  \right ) $ denote the real and imaginary parts of a complex number, respectively.

\begin{figure*}[t]
	\centering
	\includegraphics[width=6.2in]{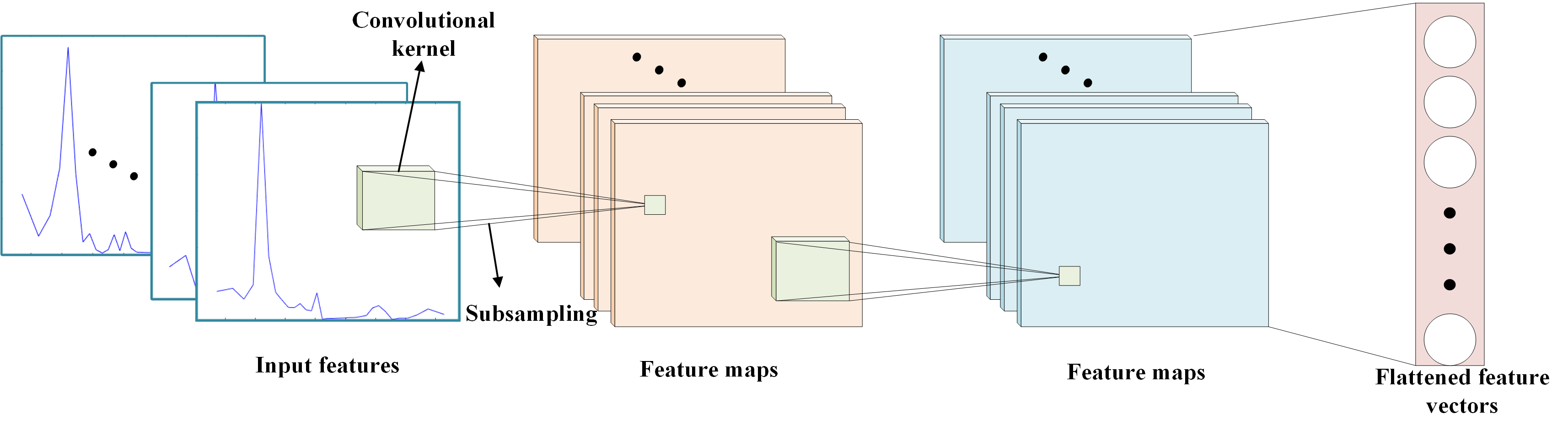}
	\vspace{-0.2cm}
	\caption{\fontsize{11pt}{\baselineskip}\selectfont Architecture of the CNN module included in the proposed NMBEnet.}
	\label{cnn}
	\vspace{-0.6cm}
\end{figure*}
$2)$ $\textit{CNN-based Feature Extraction Module}$: In this module, $L_{\mathrm{C} } $ convolutional layers are employed to extract hidden features in the input image as well as to explore the connections between different layers of the image, which is shown in Fig. \ref{cnn}.  Each convolutional layer is followed by a ReLU layer and a BatchNorm layer, whose roles are to provide nonlinear fitting capability and speed up convergence, respectively. The module ends with a flattening layer, which is responsible for flattening the extracted feature matrices of multiple layers into the vector $\mathbf{v} _{u}^{0} $. Thanks to the fact that the number of layers of the convolutional kernel is the same as that of the input image, the convolutional network can extract features from each layer of the image simultaneously, which makes it possible to explore the correlation between subcarriers. Note that the output $\left \{ \mathbf{S}_{u} \right \} _{u=1}^{U} $ from the preprocessing module is processed by $U$ same CNN modules simultaneously, generating feature vectors $\left \{ \mathbf{v} _{u}^{0}  \right \} _{u=1}^{U} $ which are then fed to the GNN module.
\begin{figure}[t]
	\centering
	\includegraphics[width=3.5in]{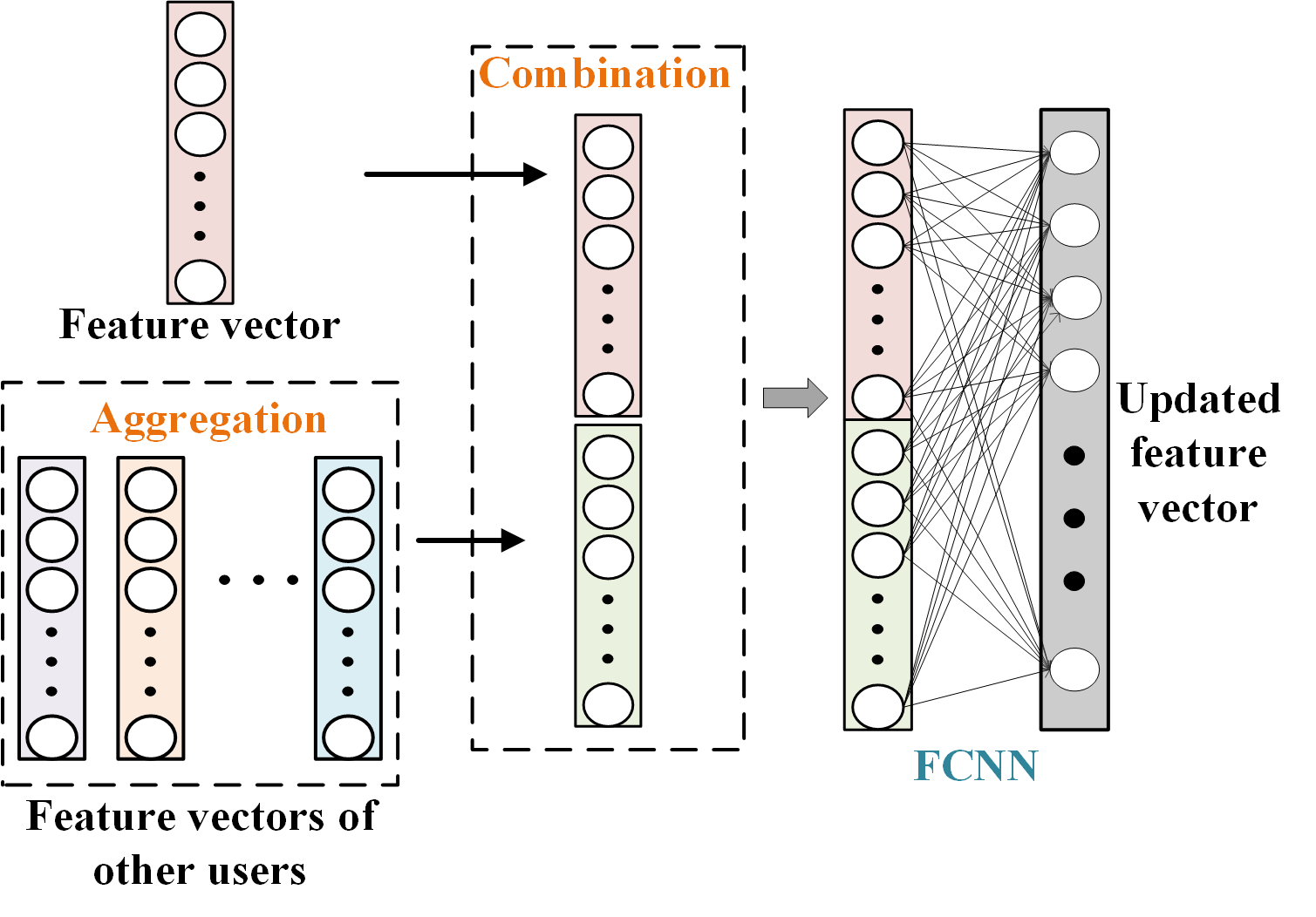}
	\vspace{-0.2cm}
	\caption{\fontsize{11pt}{\baselineskip}\selectfont Architecture of the each GNN layer included in the proposed NMBEnet.}
	\label{gnn}
	\vspace{-0.6cm}
\end{figure}

$3)$ $\textit{ GNN-based Feature Updating Module}$: After obtaining the feature vectors extracted by the CNN module from each user, we construct them into a graph-based structure, which is shown in Fig. \ref{NMBEnet}. In the constructed graph, each point represents a user and each user is characterised by its feature vector. Due to the similarity of wireless propagation conditions among users, channels between any two users are correlated. Hence, it is assumed that any two users are connected to each other, i.e., there are edges between any two nodes, where the edge means that two nodes are correlated. Based on the constructed graph, we employ $L_{\mathrm{G} } $ GNN layers to update the feature vector of each user by exploring the correlation between users. The complete structure of each GNN layer is shown in Fig. \ref{gnn}. The main part of each GNN layer is a fully connected layer. However, unlike FCNNs, GNNs add ``combination" and ``aggregation" structures in front of each fully connected layer, which help the GNNs to view and take into account the feature vectors of other users when updating the feature vector. The feature update processing for each layer can be formulated as
\begin{equation}\label{gnn_function}
	\begin{aligned}
		\mathbf{v}_{u}^{l}=f_{\mathrm{G}}^{l}\left( f_{\mathrm{com}}\left( \mathbf{v}_{u}^{l-1},f_{\mathrm{agg}}\left( \left\{ \mathbf{v}_{i}^{l-1} \right\} _{i\in \mathcal{O} _{(u)}} \right) \right) \right),
	\end{aligned}
\end{equation}
where $\mathbf{v}_{u}^{l}$ represents the output vector of the $l$-th GNN layer and $\mathcal{O} _{(u)}=\left \{ 1,2,\dots ,u-1,u+1,\dots ,U \right \}  $ denotes the indices of other users. $f_{\mathrm{G}}^{l}\left ( \cdot  \right )$ represents the mapping function of the $l$-th fully connected layer. $f_{\mathrm{agg}}\left ( \cdot  \right ) $ denotes the aggregation function, which averages the input vectors by element. For example, when the input is $\mathbf{c} _{1} ,\mathbf{c} _{2},\dots , \mathbf{c} _{n} $, the $i$-th element of the output of  $f_{\mathrm{agg}}\left ( \mathbf{c} _{1} ,\mathbf{c} _{1},\dots , \mathbf{c} _{n}  \right )  $ is given by
\begin{equation}\label{agg_function}
	\begin{aligned}
	\left [ f_{\mathrm{agg}}\left ( \mathbf{c} _{1} ,\mathbf{c} _{2},\dots , \mathbf{c} _{n}  \right ) \right ]_{i}= \mathrm{mean  }\left (\left [ \mathbf{c} _{1} \right ]_{i} ,\left [ \mathbf{c} _{2} \right ]_{i} ,\dots ,\left [ \mathbf{c} _{n} \right ]_{i}    \right )  ,
	\end{aligned}
\end{equation}
where $ \mathrm{mean  }$ denotes the averaging operation. Additionally, $f_{\mathrm{com}}\left ( \cdot  \right ) $ denotes the combination function, which splices the two input vectors into a new vector. For example, when the input is $\mathbf{c} _{1} ,\mathbf{c} _{2} $, the  output of  $f_{\mathrm{com}}\left ( \mathbf{c} _{1} ,\mathbf{c} _{1}\right )  $ is given by
\begin{equation}\label{com_function}
	\begin{aligned}
f_{\mathrm{com}}\left ( \mathbf{c} _{1} ,\mathbf{c} _{1}\right ) =\left [ \mathbf{c} _{1};\mathbf{c} _{2} \right ] .
	\end{aligned}
\end{equation}

Based on the ``combination" and ``aggregation" structures, the information of other users is also utilized in the updating of the features ultimately affecting the estimation of the optimal beam, in which the correlation between the users is explored. 

$4)$ $\textit{Output Module}$: In this module, $L_{\mathrm{F} } $ fully connected layers and a Softmax layer are employed to further process the updated feature vectors from the GNN module and map them into probability vectors about the codewords. 

Since near-field mmWave codewords require not only angle search but also distance search, we design a dual network structure which has been shown in our previous work to be effective in estimating the optimal codeword\cite{self_1}. Specifically, we construct two NMBEnets of the same structure, called angele NMBEnet and distance NMBEnet, both of which contain the four modules outlined above. Based on the pilot signals, one of the NMBEnets is trained to estimate the angle index, and the other is used to estimate the distance index of the optimal near-field mmWave codeword. Based on the dual-network structure, the pilot signals are fed into each of the two NMBEnets and two probability vectors are output, which is given by
\begin{equation}\label{prob_vector}
	\begin{gathered}
			\hat{\mathbf{p}}_{u}^{\textrm{a}}=\left [ \hat{p}_{u,1}^{\textrm{a}},\hat{p}_{u,2}^{\textrm{a}},\cdots ,\hat{p}_{u,M}^{\textrm{a}}\right ]^{T},\\
		\hat{\mathbf{p}}_{u}^{\textrm{d}}=\left [ \hat{p}_{u,1}^{\textrm{d}},\hat{p}_{u,2}^{\textrm{d}},\cdots ,\hat{p}_{u,S}^{\textrm{d}}\right ]^{T},
	\end{gathered}
\end{equation}
where $\hat{p}_{u,m}^{\textrm{a}}$ denotes the probability that the angle index of the optimal codeword for user $u$ is $m$, $\hat{p}_{u,s}^{\textrm{d}}$ denotes the probability that the distance index of the optimal codeword for user $u$ is $s$. Then, the probability that the near-field mmWave codeword $\mathbf{b}\left( \psi  _m,r_{s,m} \right)$ is the optimal codeword is given by
\begin{equation}\label{prob_code}
	\begin{aligned}
		\hat{p}_{u}^{\textrm{(n,s)}}=\hat{p}_{u,m}^{\textrm{a}}\hat{p}_{u,s}^{\textrm{d}}.
	\end{aligned}
\end{equation}

Eventually, each user's sub-6 GHz pilot signals are processed by the two NMBEnet's to be mapped into probability vector about the codeword, which is wtitten as 
\begin{equation}\label{prob_vector_code}
	\begin{aligned}
		\hat{\mathbf{p} }_{u}=\left [ \hat{p}_{u}^{\textrm{(1,1)}},\hat{p}_{u}^{\textrm{(1,2)}},\cdots,\hat{p}_{u}^{(S,M)} \right ].
	\end{aligned}
	\vspace{-0.2cm}
\end{equation}

\subsection{NMBEnet-based Near-Field mmWave Beam Training}
This subsection presents the entire NMBEnet-based beam training process, primarily comprising online and offline phases.

For the offline training phase, the training dataset is constructed to train the proposed NMBEnets, where each training data sample contains the uplink sub-6 GHz pilot signals as inputs, as well as the angle index and distance index of the optimal near-field mmWave codeword as labels. The optimal near-field mmWave beams for each user, i.e. the labels in the dataset, can be obtained by employing a traditional exhaustive beam search scheme. Based on the constructed training dataset, we train the two NMBEnets by employing the cross-loss functions, which are represented as
\begin{equation}\label{loss_fun}
	\begin{aligned}
		&\mathrm{Loss} ^{a} = - {\textstyle \sum_{u=1}^{U}} {\textstyle \sum_{m=1}^{M}}  p_{u,m}^{\textrm{a}} \mathrm{log} _{10} \hat{p}_{u,m}^{\textrm{a}},\\
		&\mathrm{Loss} ^{d} = - {\textstyle \sum_{u=1}^{U}} {\textstyle \sum_{s=1}^{S}}  p_{u,s}^{\textrm{d}} \mathrm{log} _{10} \hat{p}_{u,s}^{\textrm{d}},
	\end{aligned}
	\vspace{-0.2cm}
\end{equation}
where $ p_{u,m}^{\textrm{a}}=1$ and $ p_{u,s}^{\textrm{d}}=1$ if the optimal near-field mmWave codeword is $\mathbf{b}\left( \psi  _m,r_{s,m} \right)$, otherwise $ p_{u,m}^{\textrm{a}}=0$ and $ p_{u,s}^{\textrm{d}}=0$.

Once the two NMBEnets are fully trained, they are switched to the online estimation phase. 
Firstly, the users transmit time-orthogonal sub-6 GHz pilot signals in the uplink, and then the received signals of each user on each low-frequency subcarrier can be obtained via (\ref{sub6g_ul_r}), which is given by
\begin{equation}\label{sub_r_set}
	\begin{aligned}
	\mathcal{Y} _{sub}=\left \{\left \{  \overline{\mathbf{y}}_1[\overline{k}]  \right \}_{\overline{k}=1}^{\overline{K}} ,\dots , \left \{  \overline{\mathbf{y}}_U[\overline{k}]  \right \}_{\overline{k}=1}^{\overline{K}}   \right \}  .
	\end{aligned}
	\vspace{-0.2cm}
\end{equation}

Then, the received pilot signals are input into the two proposed NMBEnets, and the probability vectors about the angle and distance are obtained through (\ref{prob_vector}). After that, The probability vector $\left \{ \hat{\mathbf{p} }_{u} \right \} _{u=1}^{U} $ is obtained through (\ref{prob_code}) and (\ref{prob_vector_code}), which contains the probability of each codeword being the optimal codeword. Based on $\left \{ \hat{\mathbf{p} }_{u} \right \} _{u=1}^{U} $, we can obtain the optimal near-field mmWave codeword  $\mathbf{b} _{u}^{\star } =\mathbf{b}\left( \psi  _{m^{\star } } ,r_{{s^{\star } },{m^{\star } }} \right)$ for the $u$-th user, where the optimal angle index and distance index are given by
\begin{equation}\label{optimal_index}
	\begin{aligned}
	{s^{\star } },{m^{\star } }=\mathop{\arg\max}\limits_{s, m }\hat{p}_{u}^{\textrm{(s,m)}}.
	\end{aligned}
	\vspace{-0.2cm}
\end{equation}

After obtaining the optimal near-field mmWave codeword for each user $\left\{ \mathbf{b} _{1}^{\star },\mathbf{b} _{2}^{\star },\dots ,\mathbf{b} _{U}^{\star }  \right \}$, the analog precoder of the mmWave base station can be designed as
\begin{equation}\label{designed_frf}
	\begin{aligned}
		\mathbf{{f} }_u^{\textrm{RF}}=\mathbf{b} _{u}^{\star },
			\mathbf{{F} }_{\textrm{RF}}=\left [  		\mathbf{{f} }_1^{\textrm{RF}}, 	\mathbf	{{f} }_2^{\textrm{RF}} ,\dots ,		\mathbf{{f} }_U^{\textrm{RF}}\right ].
	\end{aligned}
	\vspace{-0.2cm}
\end{equation} 

When the analog precoder is determined, the equivalent channel can be estimated by sending the mmWave pilot signal uplink through the user and the digital precoder can be obtained by the ZF based precoding scheme, which is described in detail in \cite{sun} and \cite{hybrid_precoding_AA_1}.

%


\vspace{-0.3cm}

\section{Simulation Results}\label{simulation}

In this section, the performance of our proposed NMBEnet-based near-field mmWave beam training scheme using information from the sub-6 GHz band is evaluated by performing extensive simulation experiments in which the state-of-the-art software called WI is adopted.

\begin{table}[]
	\centering 
	\caption{\text { SYSTEM PARAMETERS }} 
	\label{system_table} 
	\resizebox{0.8\columnwidth}{!}{
	\begin{tabular}{lll}
		\hline
		System Parameters                                                               & sub-6 GHz  & mmWave  \\ \hline
		Center frequency $\overline{f}_{\mathrm{c}}$, ${f}_{\mathrm{c}}$       & 5.5 GHz & 30 GHz  \\
		Subcarrier number $\overline{K}$, $K$                                           & 32      & 128     \\
		Bandwidth $\overline{W}$, $W$                                                   & 10 MHz  & 10 MHz  \\
		Antenna number $\overline{M}$, $M$                                              & 32      & 256     \\
		Transmit power $\overline{P} _{\mathrm{ul} }$, ${P} _{\mathrm{dl} }$            & -10 dBm  & 2 dBm   \\
		Nosie power $\overline{\sigma }_{\mathrm{ul}}^{2}$, $\sigma _{\mathrm{dl}}^{2}$ & -81 dBm & -81 dBm \\ \hline
	\end{tabular}}
\vspace{-0.4cm}
\end{table}

\begin{table*}[]
	\centering 
	\caption{\text { NMBEnet PARAMETERS }} 
	\label{netwrok_table} 
	\resizebox{1.8\columnwidth}{!}{
		\begin{tabular}{cccc}
			\hline
			Module                                     & Network Layer & Structures                                                                              & Nimbers of parameters \\ \hline
			\multirow{3}{*}{Feature extraction module} & Convolution   & $C_{\mathrm{i} }=32$, $C_{\mathrm{0} }=64$, $C_{\mathrm{k} }=\left \{ 1,3 \right \} $, ReLU, BatchNorm    & $6.14\times10^{3}$                     \\
			& Convolution   & $C_{\mathrm{i} }=64$, $C_{\mathrm{0} }=256$, $C_{\mathrm{k} }=\left \{ 1,3 \right \} $, ReLU, BatchNorm   & $4.91\times10^{4}$                     \\
			& Flatten       &                                                                                         &                       \\ \hline
			\multirow{2}{*}{Feature updating modul}    & FC            & Combination, Aggregation, $F_{\mathrm{i} }=2304$, $F_{\mathrm{o}}=512$, ReLU, BatchNorm & $1.17\times10^{6}$                      \\
			& FC            & Combination, Aggregation, $F_{\mathrm{i} }=512$, $F_{\mathrm{o}}=512$, ReLU, BatchNorm  & $2.62\times10^{5}$                      \\ \hline
			\multirow{2}{*}{Output module}             & FC            & $F_{\mathrm{i} }=512$, $F_{\mathrm{o}}=128$, ReLU, BatchNorm                            & $6.55\times10^{4}$                      \\
			& FC            & $F_{\mathrm{i} }=128$, $F_{\mathrm{o}}=5/256$, Softmax                                  & 640 /  $3.27\times10^{4}$                    \\ \hline
	\end{tabular}}
\vspace{-0.5cm}
\end{table*}

\vspace{-0.2cm}
\subsection{System Setup}
\begin{figure}[t]
	\centering
	\includegraphics[width=2.5in]{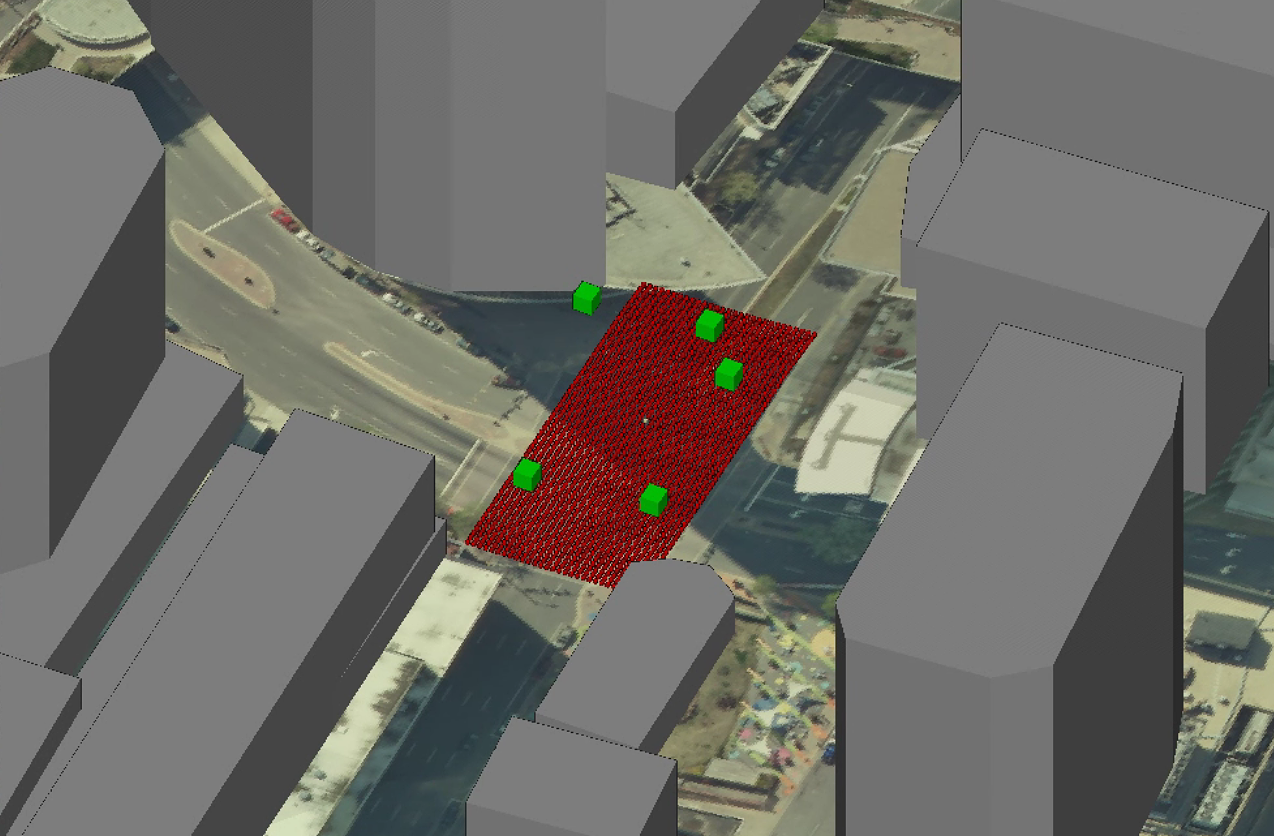}
	\vspace{-0.1cm}
	\caption{\fontsize{11pt}{\baselineskip}\selectfont Top view of the ``MIMO Example" scenario \cite{WI}}
	\label{scenario_1}
	\vspace{-0.5cm}
\end{figure}

\begin{figure}[t]
	\centering
	\includegraphics[width=2.5in]{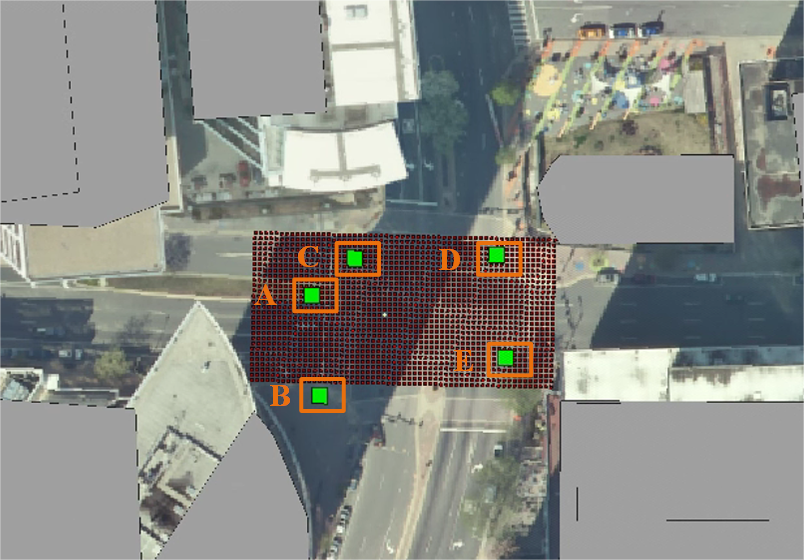}
	\vspace{-0.1cm}
	\caption{\fontsize{11pt}{\baselineskip}\selectfont Flat display of the ``MIMO Example" scenario \cite{WI}}
	\label{scenario_2}
	\vspace{-0.5cm}
\end{figure}

The WI software employs ray tracing, a mathematical technique that models signal paths from transmitter to receiver as rays, accounting for interactions with surrounding surfaces. Consequently, the channel data produced by the software encapsulates details of the wireless propagation environment. Utilizing WI simulation software enhances the assessment of our proposed beam training scheme's performance. Specifically, the ``MIMO Example" scenario that is included in the WI software is adopted and shown in Fig. \ref{scenario_1} and Fig. \ref{scenario_2}, where the mmWave BS and the sub-6 GHz base station are co-located at spot ``A" and each red spot represents a user. Furthermore, unless otherwise stated, the parameters of the mmWave system and the sub-6 GHz system are shown in Table \ref{system_table}. Based on the system parameters in Table \ref{system_table}, we set the number of sampling distances and the number of sampling angles in the near-field mmWave codebook $\mathcal{N} $ to $S=5$ and $M=256$, respectively. Thus, $MS=1280$ is the total codeword count in the near-field mmWave codebook.

Furthermore, the structure as well as the parameters of our proposed NMBEnet are shown in Table \ref{netwrok_table}, where $C_{\mathrm{i} }$, $C_{\mathrm{o} }$ and $C_{\mathrm{k} }$ denote the number of input channels, the number of output channels and the size of the convolutional kernel in the convolutional layer, respectively. $F_{\mathrm{i} }$ and $F_{\mathrm{0} }$ denote the input dimension and the output dimension in the fully-connected layer, respectively.  In the training process, the Adam optimizer is employed. We generate a dataset of 20,000 samples for training, with 95\% used for training and 5\% for validation. Training employ a learning rate of 0.006, with a decay by half after two epochs without significant accuracy improvement. The training duration spans 50 epochs, with a batch size of 800 per epoch.

\subsection{Metrics and Baselines}

Three metrics are adopted to evaluate our proposed beam training scheme and baseline schemes.

1) The sum of downlink rate on average per subcarrier is given by
\begin{equation}\label{criteria_1}
	\setlength\abovedisplayskip{3pt}
	\setlength\belowdisplayskip{3pt}
	R_\mathrm{sum}=\left (  {\textstyle \sum_{u=1}^{U}}  {\textstyle \sum_{k=1}^{K}}R_u\left[ k \right]\right )/K  .
\end{equation}

2) The effective sum rate is given by 
\begin{equation}\label{criteria_2}
	\setlength\abovedisplayskip{3pt}
	\setlength\belowdisplayskip{3pt}
	R_{\mathrm{eff}}=\left ( 1-\frac{T_{\mathrm{p}} }{T_{\mathrm{t}} }  \right ) R_{\mathrm{sum}} ,
\end{equation}
where $T_{\mathrm{t}}$ represents the overall duration of a communication session, while $T_{\mathrm{p}}$ specifically denotes the time allocated for transmitting pilot signals during this session. Conventionally, $T_{\mathrm{p}}$ is computed as the product of the number of uplink pilot symbols and the time required for transmitting each individual pilot symbol. In our simulations, we assume a transmission time of 0.1 ms per pilot symbol (or time slot) and set $T_{\mathrm{t}}$ to 0.2 $\mathrm{\mu s}$. The effective sum rate encompasses both the sum rate and the pilot overheads, underscoring the importance of achieving a high sum rate relative to pilot overhead for desirable overall performance\cite{make1}.

3) The estimation accuracy of the neural networks  $A_{cc}$ is given by
\begin{equation}\label{criteria_3}
	\setlength\abovedisplayskip{3pt}
	\setlength\belowdisplayskip{3pt}
	A_{\mathrm{cc}}=\frac{K_{\mathrm{r}} }{K} .
\end{equation}
In equation (\ref{optimal_index}), $ ({s^{\star } },{m^{\star } })$ represents the index of the codeword with the highest probability in $\hat{\mathbf{p} }_{u}$. If this index also corresponds to the optimal near-field codeword of user $k$, then user $k$ is accurately estimated by the neural network; otherwise, the estimation is inaccurate. $K{r}$ signifies the count of accurately estimated users by the neural network.

We compare the proposed NMBEnet-based near-field mmWave beam training scheme with three existing prominent beam training schemes.

1) $Baseline$ $1$: The exhaustive search scheme is adopted as Baseline 1, which requires to test all near-field mmWave codewords. The exhaustive search scheme requires the most pilot overhead but yields the best performance, so it can be used as the upper bound on the performance of beam training\cite{make1}.

2) $Baseline$ $2$ : The deep neural network model based on the FCNN structure proposed in \cite{mm_sub6g}, which utilizes sub-6 GHz channel information to predict optimal mmWave beam.

3) $Baseline$ $3$: The sub-6 GHz feature extraction model based on the CNN structure proposed in \cite{mm_sub6g}, which utilizes sub-6 GHz channel information to predict optimal mmWave beam.

\vspace{-0.3cm}
\subsection{Simulation Results}


\begin{figure}[t]
	\centering
	\includegraphics[width=2.6in]{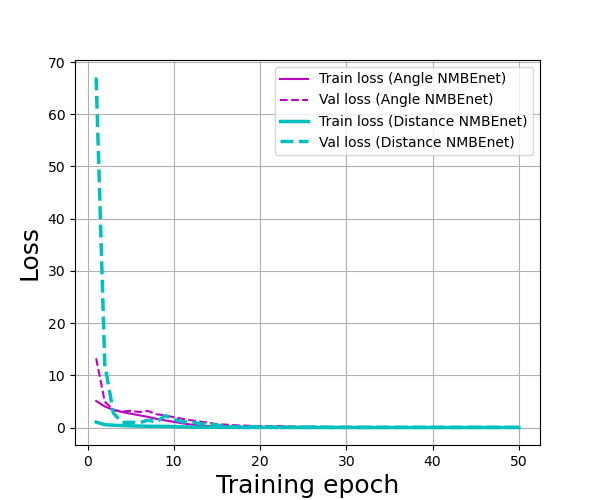}
	\vspace{-0.2cm}
	\caption{\fontsize{11pt}{\baselineskip}\selectfont Loss v.s. Training epoch}
	\label{loss}
	\vspace{-0.4cm}
\end{figure}

\begin{figure}[t]
	\centering
	\includegraphics[width=2.6in]{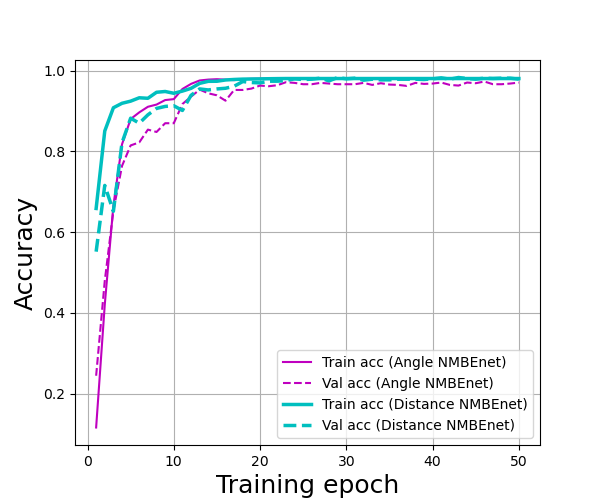}
	\vspace{-0.2cm}
	\caption{\fontsize{11pt}{\baselineskip}\selectfont Accuracy v.s. Training epoch}
	\label{acc}
	\vspace{-0.4cm}
\end{figure}

Fig. \ref{loss} and Fig. \ref{acc} show the variation of loss and accuracy with epoch during the training process of the two NMBEnets, respectively. Curves labelled with ``Train" indicate the performance of NMBEnet on the training dataset, while curves labelled with ``Val" indicate the performance of NMBEnet on the validation dataset. As can be seen in Fig. \ref{loss}, when the training rounds reach about 15, the loss functions tend to level off and the loss functions on the training dataset gradually approach that based on the validation data. At the same time, the accuracy functions reach their peaks and tend to level off. Fig. \ref{loss} and Fig. \ref{acc} illustrate that the hyperparameters we set during training, such as learning rate, batch size, etc., are suitable, which allow the angle NMBEnet and the distance NMBEnet to be adequately trained, avoiding both overfitting and underfitting.

\begin{figure*}
	\begin{minipage}[t]{0.49\linewidth}
		\centering
		\includegraphics[width=3.2in]{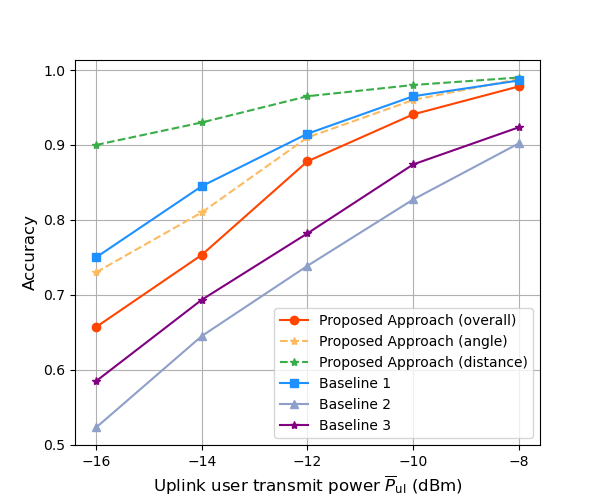}
		\vspace{-0.2cm}
		\caption{\fontsize{11pt}{\baselineskip}\selectfont Estimation accuracy v.s. uplink user transmit power. }
		\label{up_acc}
	\end{minipage}%
	\hfill
	\begin{minipage}[t]{0.49\linewidth}
		\centering
		\includegraphics[width=3.2 in]{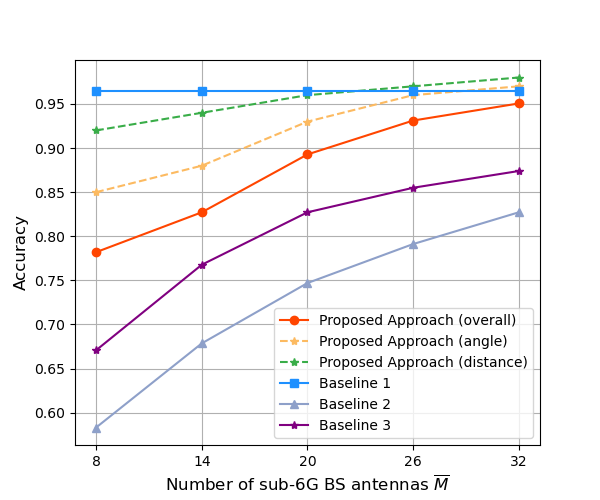}
		\vspace{-0.2cm}
		\caption{\fontsize{11pt}{\baselineskip}\selectfont Estimation accuracy v.s. the number of sub-6 GHz BS antennas. }
		\label{ant_acc}
	\end{minipage}%
	\hfill
	\vspace{-0.5cm}
\end{figure*}

\begin{figure*}
	\vspace{-0.1cm}
	\begin{minipage}[t]{0.45\linewidth}
		\centering
		\includegraphics[width=3.0in]{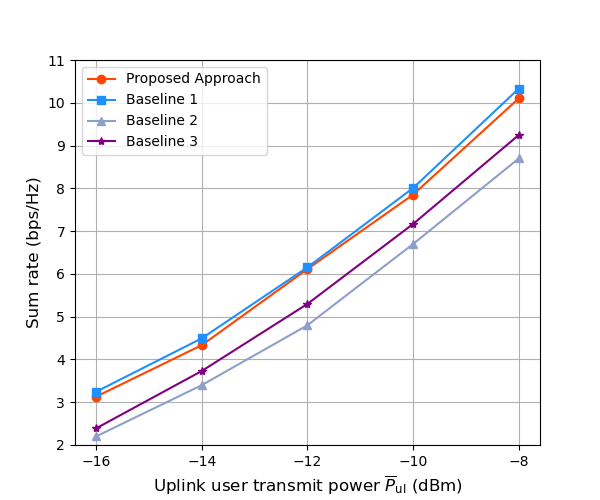}
		\vspace{-0.2cm}
		\caption{\fontsize{11pt}{\baselineskip}\selectfont Sum rate v.s. the uplink user transmit power. }
		\label{up_sum}\vspace{-0.5cm}
	\end{minipage}%
	\hfill
	\begin{minipage}[t]{0.45\linewidth}
		\centering
		\includegraphics[width=3.35in]{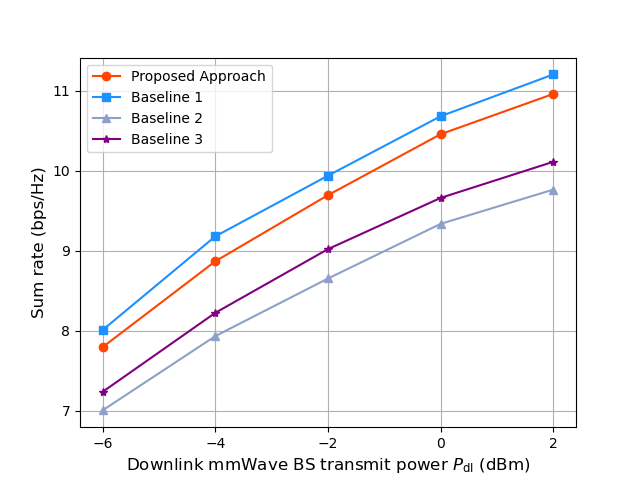}
		\vspace{-0.5cm}
		\caption{\fontsize{11pt}{\baselineskip}\selectfont Sum rate v.s. the downlink mmWave BS transmit power.}
		\label{dp_sum}\vspace{-0.5cm}
	\end{minipage}%
	\hfill
	\vspace{-0.2cm}
\end{figure*}

In Fig. \ref{up_acc}, we investigate the impact of uplink user transmit power on the accuracy of our proposed scheme and the three baseline schemes. The curve with ``angle" in the label and the curve with ``distance" in the label indicate the estimation accuracy of angle NMBEnet and distance NMBEnet, respectively. Then, the final estimation accuracy of our proposed dual NMBEnet-based beam training scheme is indicated by the curve labelled with ``overall". It can be seen that the estimation accuracies of both the proposed distance NMBEnet and angle NMBEnet increase with the uplink transmit power, which leads to the increase of the overall estimation accuracy. In addition, the estimation accuracy of angle NMBEnet is generally lower than that of distance NMBEnet, because the number of angle indices is much larger than that of distance indices. First, compared with the Baseline 2 and Baseline 3 schemes that are also based on deep learning, our proposed scheme surpasses both Baseline 2 and Baseline 3 schemes in estimating the optimal codeword across all uplink user transmission powers. This is attributed to the advantages of our proposed novel NMBEnet in exploiting inter-subcarrier and inter-user correlations. Note that the CNN and FCNN adopted in Baseline 2 and 3 are generally considered to be excellent mapping networks, but struggle to cope with beam training in OFDM multiuser scenarios. Secondly, our proposed scheme can approximate Baseline 1 in terms of accuracy, especially when $\overline{P} _{\mathrm{ul} }$ is larger than -12 dBm. It is to be noted that Baseline 1 based on exhaustive search is generally considered to be the best beam training scheme.

In Fig. \ref{ant_acc}, the impact of the number of sub-6 GHz BS's antennas on various beam training schemes is investigated. Again, it can be seen from Fig. 4 that our proposed scheme outperforms Baseline 2 and the Baseline 3 under any number of antennas. As the number of sub-6 GHz BS's antennas increases, more information about the sub-6 GHz band becomes available, which contributes to the estimation of the optimal near-field mmWave codeword for the proposed scheme and Baselines 2, 3. Hence, it can be seen from Fig. \ref{ant_acc} that the accuracy of both the proposed scheme and Baselines 2, and 3 increases with the number of antennas. However, Baseline 1 is not affected by the number of sub-6 GHz BS's antennas, because Baseline 1 does not utilize the information about the sub-6 GHz band.

Taking Fig. \ref{up_acc} and Fig. \ref{ant_acc} together, it can be seen that the proposed NMBEnet-based near-field mmWave beam training scheme can perform the dual mapping of far-field to near-field and low-frequency to high-frequency as well. It can also be seen that both angle NMBEnet and distance NMBEnet exhibit high estimation accuracy for each system parameter, which demonstrates that they can effectively extract the angle information and distance information of the optimal codeword from the pilot signals, respectively. In particular, it is important to note that the input to the neural network employed in our proposed scheme is the uplink pilot signals that can be obtained directly, rather than the system parameters or channel state information that needs to be estimated. The excellent estimation accuracy of our proposed scheme demonstrates that the uplink pilot signals can be directly used to estimate the optimal near-field mmWave codewords and proves the feasibility of the proposed end-to-end architecture.

In Fig. \ref{up_sum}, we further investigate the impact of uplink user transmit power on the sum rate. The total transmit power of the downlink mmWave base station is set to 20 dBm, i.e., $K{P} _{\mathrm{dl} }=20$ dBm. As depicted in Fig. \ref{up_sum}, our proposed scheme closely approximates the performance of Baseline 1, which employs exhaustive search, while surpassing Baselines 2 and 3, both of which are deep learning-based algorithms, in terms of downlink sum rate. Something similar is also seen in Fig. \ref{dp_sum}, which depicts the effect of downlink mmWave base station transmit power on the downlink sum rates. The exceptional estimation accuracy of our proposed scheme primarily contributes to its outstanding performance, yielding substantial gains in downlink beamforming and ultimately facilitating a high sum rate.

\begin{figure*}
	\vspace{-0.1cm}
	\begin{minipage}[t]{0.45\linewidth}
		\centering
		\includegraphics[width=3.2in]{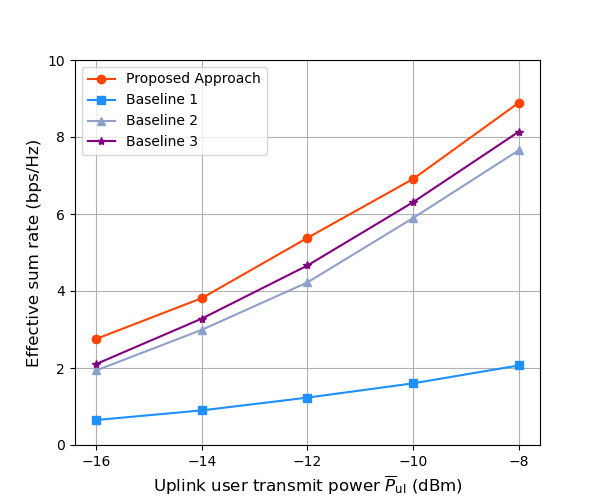}
		\vspace{-0.2cm}
		\caption{\fontsize{11pt}{\baselineskip}\selectfont Effective sum rate of different schemes as function of the uplink user transmit power.  }
		\label{up_eff_sum}\vspace{-0.5cm}
	\end{minipage}%
	\hfill
	\begin{minipage}[t]{0.45\linewidth}
		\centering
		\includegraphics[width=3.2in]{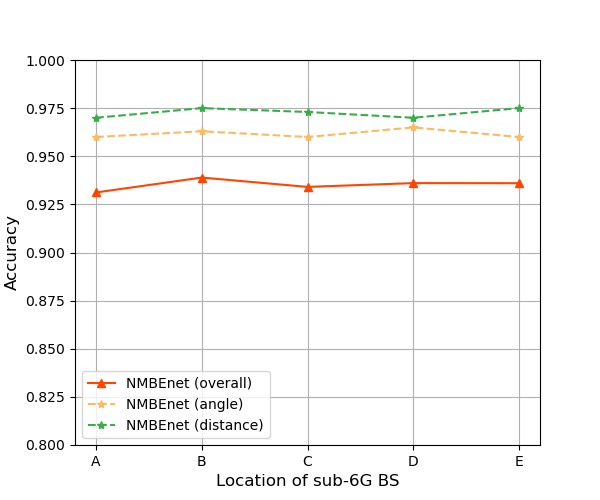}
		\vspace{-0.2cm}
		\caption{\fontsize{11pt}{\baselineskip}\selectfont Impact of sub-6 GHz BS's location on accuracy for various schemes.}
		\label{bs_acc}\vspace{-0.5cm}
	\end{minipage}%
	\hfill
	\vspace{-0.2cm}
\end{figure*}

While Baseline 1, utilizing exhaustive search, outperforms the proposed scheme in terms of data rate, it necessitates considerable pilot overheads. The extensive pilot overheads would lead to insufficient time for transmitting data during a communication session. The superiority of our proposed scheme is difficult to distinguish on the basis of sum rate metrics alone. To this end, we investigate the impact of the uplink transmit power on the effective rate, which is shown in Fig. \ref{up_eff_sum}. It can be seen from Fig. \ref{up_eff_sum} that Baseline 1 performs much worse than our proposed scheme due to the penalty it receives for excessive overhead. However, there is no performance degradation in our proposed scheme and the Baselines 2, and 3 since they only use information outside the mmWave band.

In Fig. \ref{bs_acc}, we investigate the impact of different locations of sub-6 GHz BS on the estimation accuracy of the proposed scheme. Indeed, some of the existing studies rely on the spatial similarity between sub-6 GHz BS and mmWave BS when they are located at the same position\cite{hashemi,gaofeifei}. However, it can be seen in Fig. \ref{bs_acc} that our proposed NMBEnet-based algorithm maintains high estimation accuracy at each sub-6 GHz BS's location. This shows that our proposed scheme does not depend on the spatial similarity between the BSs of the two frequency bands, which makes the proposed scheme more feasible.

\vspace{-0.4cm}

\section{Conclusions}\label{conclusion}
In this paper, we revealed and demonstrated the similarity between far-field sub-6 GHz channels and near-field mmWave channels.  Motivated by this, we proposed a deep learning-based near-field mmWave beam training scheme which exploits the information of the sub-6 GHz band to reduce the pilot overheads.  In the proposed beam training scheme, we employed the uplink sub-6 GHz guide signal to directly estimate the optimal near-field mmWave codeword. Such end-to-end design can prevent complex channel estimation. Furthermore, we proposed a novel neural network structure called NMBEnet to perform the mapping from far-field sub-6 GHz signals to optimal near-field mmWave codeword. The proposed NMBEnet mainly consist of a CNN module and a GNN module, which can improve the accuracy of beam training by leveraging inter-subcarrier correlation and inter-user correlation, respectively.
The simulation results demonstrated that the proposed scheme can obtain higher beam training accuracy than the existing deep learning-based beam training schemes.


\bibliographystyle{IEEEtran}
\vspace{-0.4cm}
\bibliography{myre}


\end{document}